%% file: main.tex
\documentclass[11pt]{article}

\date{}

\usepackage[utf8]{inputenc} 
\usepackage[T1]{fontenc}    
\usepackage{url}            
\usepackage{booktabs}       
\usepackage{amsfonts}       
\usepackage{nicefrac}       
\usepackage{microtype,nicefrac}      
\usepackage{xspace}
\usepackage{natbib}
\usepackage{comment}
\usepackage{amsmath}
\usepackage{graphicx}
\usepackage{rotating}
\usepackage{amssymb}
\usepackage{autobreak}
\usepackage{mathtools}
\usepackage[dvipsnames]{xcolor}
\usepackage{bbm}
\usepackage[ruled,vlined, linesnumbered]{algorithm2e}
\usepackage{algorithmic}
\usepackage[parfill]{parskip}

\usepackage{color}
\usepackage[urlcolor=blue,citecolor=black,linkcolor=black]{hyperref}
\usepackage{amsmath} 
\usepackage{graphicx}
\usepackage{subfig}
\usepackage{amssymb} 
\allowdisplaybreaks

\title{Deep Neural Helmholtz Operators for 3D Elastic Wave Propagation and Inversion}
\author{Caifeng Zou$^1$, Kamyar Azizzadenesheli$^2$, Zachary E. Ross$^1$, and Robert W. Clayton$^1$ \\
  $^1$ \textit{Seismological Laboratory, California Institute of Technology, Pasadena, CA \emph{91125}, USA }\\
  $^2$ \textit{Nvidia Corporation, Santa Clara, CA \emph{95051}, USA}
  }

\begin{document}

\label{firstpage}

\maketitle

\begin{abstract}
 Numerical simulations of seismic wave propagation in heterogeneous 3D media are central to investigating subsurface structures and understanding earthquake processes, yet are computationally expensive for large problems. This is particularly problematic for full waveform inversion, which typically involves numerous runs of the forward process. In machine learning there has been considerable recent work in the area of operator learning, with a new class of models called neural operators allowing for data-driven solutions to partial differential equations. Recent works in seismology have shown that when neural operators are adequately trained, they can significantly shorten the compute time for wave propagation. However, the memory required for the 3D time domain equations may be prohibitive. In this study, we show that these limitations can be overcome by solving the wave equations in the frequency domain, also known as the Helmholtz equations, since the solutions for a set of frequencies can be determined in parallel. The 3D Helmholtz neural operator is 40 times more memory-efficient than an equivalent time-domain version. We employ a U-shaped neural operator for 2D and 3D elastic wave modeling, achieving two orders of magnitude acceleration compared to a baseline spectral element method. The neural operator accurately generalizes to variable velocity structures and can be evaluated on denser input meshes than used in the training simulations. We also show that when solving for wavefields strictly on the surface, the accuracy can be significantly improved via a graph neural operator layer. In leveraging automatic differentiation, the proposed method can serve as an alternative to the adjoint-state approach for 3D full-waveform inversion, reducing the computation time by a factor of 350.
\end{abstract}


\section{Introduction}
Numerical simulations of seismic wave propagation serve as the foundation for a wide array of seismological investigations, including subsurface imaging, ground motion simulation for seismic hazard assessment, and deriving earthquake source properties. However, computational cost and memory requirements become major concerns for large 3D problems, as well as inverse problems, which require numerous evaluations of the forward model. In addition, the availability of denser seismic networks spanning wider regions offers an unprecedented wealth of data to be used. Therefore, accelerating seismic wave modeling has become a pressing need for modern seismology to align with the rapid advancements in big data and computer capabilities.

Traditional methods for solving partial differential equations (PDEs) are based on brute-force numerical schemes that discretize a physical domain and solve a governing equation on a grid. This discretization introduces a trade-off between the speed and accuracy: coarser grids provide faster results but with reduced accuracy, while finer grids offer higher accuracy at the cost of slower computations. In seismology, the finite difference method (FDM) and finite element method (FEM) are generally the most common approaches for numerical simulation of wave propagation \citep{Kelly,Olsen,Fichtner,Basabe}. As a version of FEM, the spectral element method (SEM) developed by \cite{Komatitsch} combines the higher-order accuracy from spectral methods with the flexibility of FEMs. For any of the above-mentioned solvers, a certain number of elements or grid points per seismic wavelength are required to achieve a specific level of accuracy. The most demanding calculations are at high frequencies and the slowest parts require the densest grids. The computational expense remains a bottleneck for conventional numerical methods, especially in 3D.

The recent emergence of machine learning approaches has the potential to overcome the limitations of conventional PDE solvers in computing speed and cost, especially for large inverse problems \citep{Raissi,Li2020a,Li2020b,Khoo,Lu}. To solve PDEs with neural networks, one can simply discretize the input (e.g., the elastic properties of the continuum) and output (e.g., the displacement wavefield) function spaces into finite-dimensional grids, which naturally fits into a standard neural network framework such as the convolutional neural network \citep{Guo,ZhuZabaras,Winovich,Moseley}. However, this approach is limited to a specific discretization and cannot provide solutions for new, abitrary meshes within the same physical domain. The so-called physics-informed neural networks (PINNs) \citep{Raissi} allow for the querying of solutions at new points by directly parameterizing the solution function as a neural network. PINNs penalize the residual of the governing equation in the loss function, which can be trained on their own and do not necessarily need data from a separate numerical solver. 

In the last few years, there have been a number of studies applying PINNs to seismological problems. They have been used successfully to solve Eikonal equations and derive earthquake hypocenters \citep{Smith,SmithHypoSVI}. PINNs have been used to solve the 2D and 3D scattered forms of the frequency-domain acoustic wave equations \citep{SongAlkhalifah,Alkhalifah,Song2021,huang2022pinnup,huang2022single}. Other applications lie in acoustic wave propagation \citep{moseley2020solving,Song2022}, inversion \citep{Rasht‐Behesht,zhang2023seismic}, and wavefield-modeling-based source imaging \citep{huang2023microseismic}. Compared to acoustic wave equations, the use of PINNs for  the more computationally expensive solutions to elastic wave equations is sparsely documented \citep{ren2022seismicnet,Song2023}.

However, in most realistic cases these models only learn a specific instance of the PDE and generally require training a new network for every new instance of the PDE parameters (e.g., elastic media and source properties). To address the shortcomings of the above-mentioned methods, \cite{Li2020b} proposed the discretization-invariant neural operator that learns an entire family of PDEs, removing the need for retraining with varied PDE coefficients. The universal approximation theorem \citep{Hornik} serves as the theoretical foundation for the neural operators composed of linear operators and nonlinear activations to approximate any given nonlinear continuous operator \citep{Kovachki}. 

Neural operators have seen increasing usage in seismology in recent years. The appeal of neural operators is that one can train a single deep learning model and apply it to arbitrary PDE parameter functions. Neural operators have been explored for 2D acoustic \citep{Yang2021,Li2022}, elastic \citep{Yang2023,Zhang}, and viscoelastic wave modeling \citep{Wei}, high-frequency wavefield extrapolation \citep{SongWang}, and earthquake locating \citep{Sun}. The only documented application by far in 3D seismic wave propagation was conducted by \cite{Lehmann}, who simulated the ground motions by mapping the depth of the geological model to the time of the velocity wavefields. 

Most of the applications have thus-far been limited to 2D, as the amount of GPU memory required for 3D wave propagation in the time domain can quickly end up in the range of hundreds of GB. This is because the time domain problem requires a 4D neural operator, and so the memory requirements grow extremely quickly. One way to address this issue is to solve the wave equations in the frequency domain, which are also referred to as the Helmholtz equations. In this way, the 4D machine learning model can be simplified to a 3D model. Moreover, individual frequencies can be handled independently, benefitting further from parallelized computing. The time-domain solutions can be obtained by taking the inverse Fourier transform of the frequency-domain solutions. 

In this study, our contributions are as follows. We propose a neural operator for 3D elastic wave propagation and inversion. We train a general solution operator, parameterized as a U-shaped deep neural operator, to solve the Helmholtz equations for variable velocity structures, source locations, mesh discretizations, and multiple frequencies. We provide options to model either complete wavefields or ground motions only, with the latter being more accurately predicted via a graph neural operator (GNO) layer~\citep{Li2020b}. The size of the Helmholtz neural operator is only one-fortieth of that of an equivalent 3D time domain version. The 3D Helmholtz neural operator is roughly 100 times faster than the baseline numerical solvers for forward propagation and 350 times faster for inverse problems. The trained neural operator combined with automatic differentiation facilitates rapid full-waveform inversion (FWI) with accuracy measured by relative L2 misfit of 0.03. 

\section{Methods}

\subsection{Elastic wave equations}
The time-domain elastic wave equation for inhomogeneous, isotropic media is given by 
\begin{equation}
\begin{aligned} 
 \rho \frac{\partial^2 \textbf{u}}{\partial t^2}=
&\nabla \lambda\left ( \nabla \cdot \textbf{u} \right )+\nabla \mu \cdot \left [ \nabla \textbf{u}+\left ( \nabla \textbf{u} \right )^{T} \right ]+ \\
&\left ( \lambda +2\mu  \right )\nabla\left ( \nabla\cdot \textbf{u} \right )-\mu \nabla\times \nabla\times \textbf{u}+\textbf{f},
\label{teq}
\end{aligned}
\end{equation}
where $\rho$ is the density, $\textbf{u}$ is the displacement vector, $\lambda$ and $\mu$ are the Lamé parameters, and $\textbf{f}$ is the body force (or source term). Equation 1 can be expressed in terms of P-wave velocity $V_P$ and S-wave velocity $V_S$ through
\begin{equation}
V_{P}=\sqrt{\frac{\lambda +2\mu }{\rho }}, V_{S}=\sqrt{\frac{\mu }{\rho }}.
\label{ps}
\end{equation}
The time-dependent wave equation is an initial value problem, which can be solved by explicit or implicit time-stepping schemes. The Fourier transform of the time-domain equations are referred to as the Helmholtz equations:
\begin{equation}
\begin{aligned} 
\omega ^{2}\rho U_{x}
&+\frac{\partial }{\partial x}\left [ \lambda \left ( \frac{\partial U_{x}}{\partial x}+\frac{\partial U_{y}}{\partial y} +\frac{\partial U_{z}}{\partial z}\right )+2\mu \frac{\partial U_{x}}{\partial x} \right ]\\
&+\frac{\partial }{\partial y}\left [ \mu \left ( \frac{\partial U_{x}}{\partial y} +\frac{\partial U_{y}}{\partial x}\right ) \right ]\\
&+\frac{\partial }{\partial z}\left [ \mu \left ( \frac{\partial U_{x}}{\partial z} +\frac{\partial U_{z}}{\partial x}\right ) \right ]=-F_{x},\\
\omega ^{2}\rho U_{y}
&+\frac{\partial }{\partial x}\left [ \mu \left ( \frac{\partial U_{x}}{\partial y} +\frac{\partial U_{y}}{\partial x}\right ) \right ]\\
&+\frac{\partial }{\partial y}\left [ \lambda \left ( \frac{\partial U_{x}}{\partial x}+\frac{\partial U_{y}}{\partial y} +\frac{\partial U_{z}}{\partial z}\right )+2\mu \frac{\partial U_{y}}{\partial y} \right ]\\
&+\frac{\partial }{\partial z}\left [ \mu \left ( \frac{\partial U_{y}}{\partial z} +\frac{\partial U_{z}}{\partial y}\right ) \right ]=-F_{y},\\
\omega ^{2}\rho U_{z}
&+\frac{\partial }{\partial x}\left [ \mu \left ( \frac{\partial U_{x}}{\partial z} +\frac{\partial U_{z}}{\partial x}\right ) \right ]\\
&+\frac{\partial }{\partial y}\left [ \mu \left ( \frac{\partial U_{y}}{\partial z} +\frac{\partial U_{z}}{\partial y}\right ) \right ]\\
&+\frac{\partial }{\partial z}\left [ \lambda \left ( \frac{\partial U_{x}}{\partial x}+\frac{\partial U_{y}}{\partial y} +\frac{\partial U_{z}}{\partial z}\right )+2\mu \frac{\partial U_{z}}{\partial z} \right ]=-F_{z},
\label{feq}
\end{aligned}
\end{equation}
where $\omega$ is the angular frequency, $\textbf{U}=\left ( U_{x},U_{y},U_{z} \right )$ is the displacement vector in the frequency domain, and $\textbf{F}=\left ( F_{x},F_{y},F_{z} \right )$ is the frequency-domain body force. The wave propagation in the frequency domain becomes a boundary value problem, which can be expressed in a more compact, discretized formulation \citep{Pratt}:
\begin{equation}
\begin{aligned} 
\textbf{L}\left ( \omega  \right )\left [ \textbf{U}\left ( \omega  \right ) \right ]=-\textbf{F}\left ( \omega  \right ),
\label{comp}
\end{aligned}
\end{equation}
where $\textbf{L}\left ( \omega  \right )$ is an $N\times N$ matrix and $N$ is the number of discretization points. In this linear system, the frequency-domain wavefield $\textbf{U}\left ( \omega  \right )$ is related to the source term $\textbf{F}\left ( \omega  \right )$ via a sparse matrix $\textbf{L}\left ( \omega  \right )$. The solution to Equation \ref{comp} can be obtained using either direct or iterative solvers. Direct solvers, such as lower-upper (LU) decomposition or Cholesky factorization, suffer from intensive memory requirements and prohibitive computational time for large linear systems. Iterative solvers heavily rely on a well-designed preconditioner to prevent divergence or slow convergence \citep{HuangGreen}.

\subsection{Background on Neural Operators}
Neural operators \citep{Li2020b} are a class of models composed of linear integral operators and nonlinear activations. Such models are universal approximators of arbitrary nonlinear continuous operators \citep{Kovachki}. Let $\mathcal{L}_a$ be a linear differential operator determined by parameter $a$ and consider the following PDE:
\begin{equation}
\begin{aligned} 
\left ( \mathcal{L}_a u \right )\left ( x \right )=f\left ( x \right ),x\in D,
\label{luf}
\end{aligned}
\end{equation}
where $u$ is the PDE solution, $f$ is a fixed function such as the source term in a wave equation, and $D\subset \mathbb{R}^{d}$ is a bounded, open set. The Green’s function $G$ is defined as the unique solution to
\begin{equation}
\begin{aligned} 
\mathcal{L}_{a}G\left ( x,\cdot  \right )=\delta _x,
\label{lgd}
\end{aligned}
\end{equation}
where $\delta _x$ is the Dirac delta function centered at $x$. Note that $G$ depends on $a$, so the solution to Equation \ref{luf} is given by
\begin{equation}
\begin{aligned} 
u\left ( x \right )=\int_{D}G_{a}\left ( x,y \right )f\left ( y \right )\mathrm{d}y.
\label{usol}
\end{aligned}
\end{equation}
In seismology, Equation \ref{usol} remains a linear system when the velocity structure, $a(x)$ is fixed, and describes the mapping from the input source function $f\left(x\right)$ to the output wavefield solution $u\left(x\right)$. This study considers a more complicated problem, however, in which we aim to obtain a solution operator that maps from $a(x)$ to $u(x)$. This operator becomes nonlinear and is not known in closed form; here we instead aim to approximate it with a learnable parametric model, i.e. a neural operator. These neural operators incorporate a nonlinear point-wise activation function following each linear integral operator. A neural operator containing $L$ layers is formulated with an iterative architecture in the following manner:
\begin{equation}
\begin{aligned} 
&v_{0}\left ( x \right )=\left ( \mathcal{P}\,a \right )\left ( x \right ),\\
&v_{l+1}\left ( x \right )=\sigma \left ( \mathcal{W}_{l} \,v_{l}\left ( x \right )+\int_{D}\kappa _{l} \left ( x,y \right )v_{l} \left ( y \right )\mathrm{d}y\right ),\\
&u\left ( x \right )=\left ( \mathcal{Q}\,v_{L} \right )\left ( x \right ),l=0,..,L-1,
\label{no}
\end{aligned}
\end{equation}
where $a\left(x\right)$ is the input function(s) (e.g., velocity and source functions), $u\left(x\right)$ is the output function(s) (e.g., displacement wavefields), $v_l$ is the hidden representation of the $l^{th}$ layer and is input to the next layer, $\mathcal{P}$ is a point-wise operator lifting the input to a higher dimensional representation, $\mathcal{Q}$ is a point-wise operator projecting the last hidden representation to the output dimensionality, $\mathcal{W}_l$ is a point-wise operator included to learn nonperiodic behavior on the boundaries of the domain, $\kappa_{l}$ is a parametric kernel function, and $\sigma$ is a point-wise nonlinear activation. In order to speed up computations, \cite{Li2020a} proposed the Fourier neural operator (FNO) which replaces the kernel integral operator with a convolution operator defined in the Fourier space:
\begin{equation}
\begin{aligned} 
\int_D\kappa _l\left ( x,y \right )v_{l}\left ( y \right )\mathrm{d}y=\mathcal{F}^{-1}\left ( \mathcal{F}\left ( \kappa _l \right )\cdot\mathcal{F}\left ( v_l \right ) \right ),
\label{fno}
\end{aligned}
\end{equation}
where $\mathcal{F}$ and $\mathcal{F}^{-1}$ denote the Fourier transform and inverse Fourier transform, respectively. 

\begin{figure*}
\includegraphics[width=1.0\textwidth]{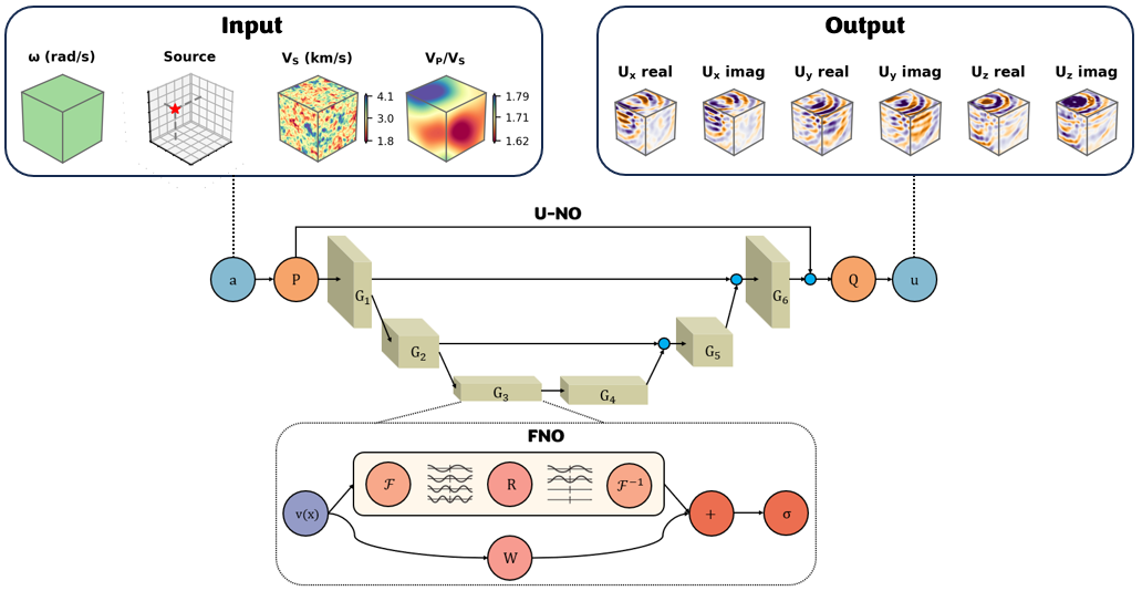}
\caption{U-NO architecture. The input a comprises P- and S- wave velocity ($V_P$ and $V_S$), the source location, and the frequency as a constant function. The output $u$ comprises the frequency-domain displacement wavefields. $\mathcal{P}$ denotes a point-wise lifting operator, $\mathcal{Q}$ denotes a point-wise projection operator, and $\mathcal{G}$ denotes an FNO as the inner integral operator. Smaller blue circles denote function-space concatenations. Inside the dotted box is the composition of each FNO layer, where $v$ is the layer input, $\mathcal{F}$ is the Fourier transform, $\mathcal{F}^{-1}$ is the inverse Fourier transform, $\mathcal{R}$ and $\mathcal{W}$ are linear operators, and $\sigma$ is the nonlinear activation function.}
\label{uno}
\end{figure*}

Since the development of FNO, other neural operator models have been developed to improve upon it further. Inspired by the U-net \citep{Ronneberger}, \cite{Rahman} designed a U-shaped neural operator (U-NO) that allows for much deeper models by progressive contraction and expansion of the physical domain and co-domain. These deeper models achieve better performance and memory usage, and are the choice used in this study. Specifically, we use a 6-layer U-NO taking the FNO as the inner integral operator, as illustrated in Fig. \ref{uno}.

Note that in this study the U-NO learns the solution in the frequency domain, mainly to break the memory bottleneck. For a 3D problem, the solution to the time-domain wave equation has a form of $\textbf{u}(x,y,z,t)$. This requires a neural operator with four dimensions, which will be significantly memory demanding. However, the solution to the frequency-domain wave equation (Helmholtz equation) can be divided by frequencies. Each individual frequency has its independent solution in the form of $\textbf{U}_{\omega}(x,y,z)$, so that a 3D neural operator is qualified. For the 3D problems in this study, an equivalent time-domain neural operator can take up 215 GB only for model parameters. But with the proposed Helmholtz neural operator, the model size can be reduced by a factor of 40. In addition to saving model memory, this formulation enables parallelization at the frequency level to accelerate model training and reduce data memory.

\subsection{Automatic Differentiation}
Automatic differentiation (AD) exploits the fact that all numerical computations can be seen as a sequence of elementary operations with known derivatives, and by applying the chain rule to the derivatives of the constituent operations we can obtain the derivative of the overall composition \citep{Griewank}. AD falls between numerical differentiation and symbolic differentiation. It differentiates common functions and expressions using a symbolic differentiation approach and populates the obtained derivatives with numerical values, saving the intermediate results. These saved results are then combined to obtain the final desired value. AD is compatible not only with closed-form expressions but also with control flow structures such as loops, branching, recursion, and procedure calls, because they will all be translated into numeric evaluation traces which form the basis of AD \citep{Baydin}.

There are two modes of automatic differentiation, the forward mode and the reverse mode. In the forward mode, derivatives of all dependent variables with respect to a certain independent variable are computed simultaneously with the forward propagation of the computational graph. Hence, the forward accumulation is more efficient in scenarios where the number of independent variables is significantly smaller than the number of dependent variables. However, in machine learning practice, the primary focus revolves around computing the gradients of a scalar-valued objective function with respect to a large number of parameters. This establishes the reverse-mode automatic differentiation as the cornerstone of the backpropagation algorithm. As opposed to the forward mode, the reverse-mode AD is a two-phase process. Firstly, a forward pass of the code is implemented to populate the intermediate variables and record the dependencies in the computational graph. Secondly, gradients are computed by propagating the derivatives back from the objective function to the parameters of interest using the chain rule. These gradients are later employed by a gradient descent method to iteratively update the parameters of interest, aiming to minimize the objective function.

\section{Data}
In this study, we train neural operators with supervised learning, i.e. a set of previously computed numerical simulations are used to learn from. We use a SEM software package named SALVUS \citep{Afanasiev} to generate 31000 simulations for both 2D and 3D data sets, hereafter referred to as the ground truth. Within each simulation data set, 27000 simulations are used for neural operator training, 3000 simulations are used to validate the model hyperparameters, and 1000 simulations are used to test the generalizability of the trained models. To further demonstrate the generalization ability, a 3D overthrust velocity model depicting complex thrusting overlying an extensional and rift sequence \citep{Aminzadeh} is used for U-NO-based waveform modeling and full-waveform inversion. 

For the SEM solver, the computational domain is 5 km × 5 km (× 5 km). We take one element per wavelength at the maximum frequency and a polynomial degree of 4, as recommended in the SALVUS documentation. This combination should provide sufficient grid points to sample the domain. The time step for numerical simulation is set to 0.002 s to satisfy the Courant–Friedrichs–Lewy (CFL) \citep{Courant} condition. The displacement wavefield is excited with a Ricker wavelet with a central frequency of 3 Hz. The spatial component of this source is a delta function, which in practice is approximated by a narrow Gaussian for differentiability. The Ricker source is configured as an isotropic explosive source randomly placed within the computational domain. The S-wave velocity ($V_S$) models are generated from a background of 3 km/s perturbed by von Kármán-type random fields \citep{von} with a Hurst exponent of 0.5, correlation length of 8 grid cells, and a fractional magnitude of the fluctuation set to 10\%. Several authors have shown that the von Kármán correlation function can represent the Earth’s inhomogeneities\citep{Chemingui,Mai,Nakata}. The P- to S-wave velocity ratio ($V_P/V_S$) models are generated from a background of 1.732 perturbed by Gaussian random fields with a correlation length of 32 grid cells and standard deviation of 2\%. The P-wave velocity ($V_P$) models are obtained by multiplying $V_S$ by $V_P/V_S$ and the density is set to a constant 2.7 g/cm³. The free-surface condition is set for the top boundary, while absorbing boundary conditions \citep{Clayton} are configured for the other five sides in 3D or three edges in 2D. Displacement wavefields with a total duration of 2.5 s are simulated in both 2D and 3D. Each 2D simulation takes about 1.6 s using one GPU with a memory usage of 0.3 GB and each 3D simulation takes about 30 s using one GPU with a memory usage of 0.8 GB.

The source-time functions and solutions to the elastic wave equations from SALVUS are Fourier transformed for use with U-NO. As our model computes a solution for a single frequency component, the number of training samples available for learning is the number of simulations in time multiplied by the number of frequencies of interest. Alternatively, if a Helmholtz solver is available, it would be possible to directly generate solutions for a desired frequency and avoid this extra computational overhead. The input to U-NO consists of only spatial functions, comprising $V_P$, $V_S$, the complex-valued source term indicating the source location, and the frequency desired given as a spatially-constant function. The output of U-NO is either a 2D or 3D complex-valued displacement wavefield that corresponds to the given frequency input. By evaluating the forward model for multiple frequencies, we can return to the time domain by taking the inverse Fourier transform. In this formulation, frequencies of interest can be tackled by U-NO in parallel, making the problem highly efficient for use with GPUs. Both the input and output data are standardized using the statistics of training data for basic pre-processing.

Note that neural operators are not required to adhere to the spatial and temporal discretization schemes of the numerical solver. For computational efficiency, the time step for U-NO is set to 0.05 s, with the Nyquist-Shannon sampling theorem taken into account. The spatial discretization adopts a regular mesh of 64 × 64 (× 64) covering the same physical domain as the SEM solver. To improve computational efficiency further, frequencies with little energy in the power density spectrum of the wavefield are eliminated. The significant frequencies range from 1.6 Hz to 6 Hz with an interval of 0.4 Hz determined by the 2.5-s duration, which are able to reconstruct the time-domain waveform. Thus by evaluating the solution in the Fourier domain, we can gain performance advantages by not computing solutions for unnecessary frequencies.

\section{Results}
\subsection{2D elastic wave modeling with U-NO}
\begin{figure}
\centering
\includegraphics[width=0.7\textwidth]{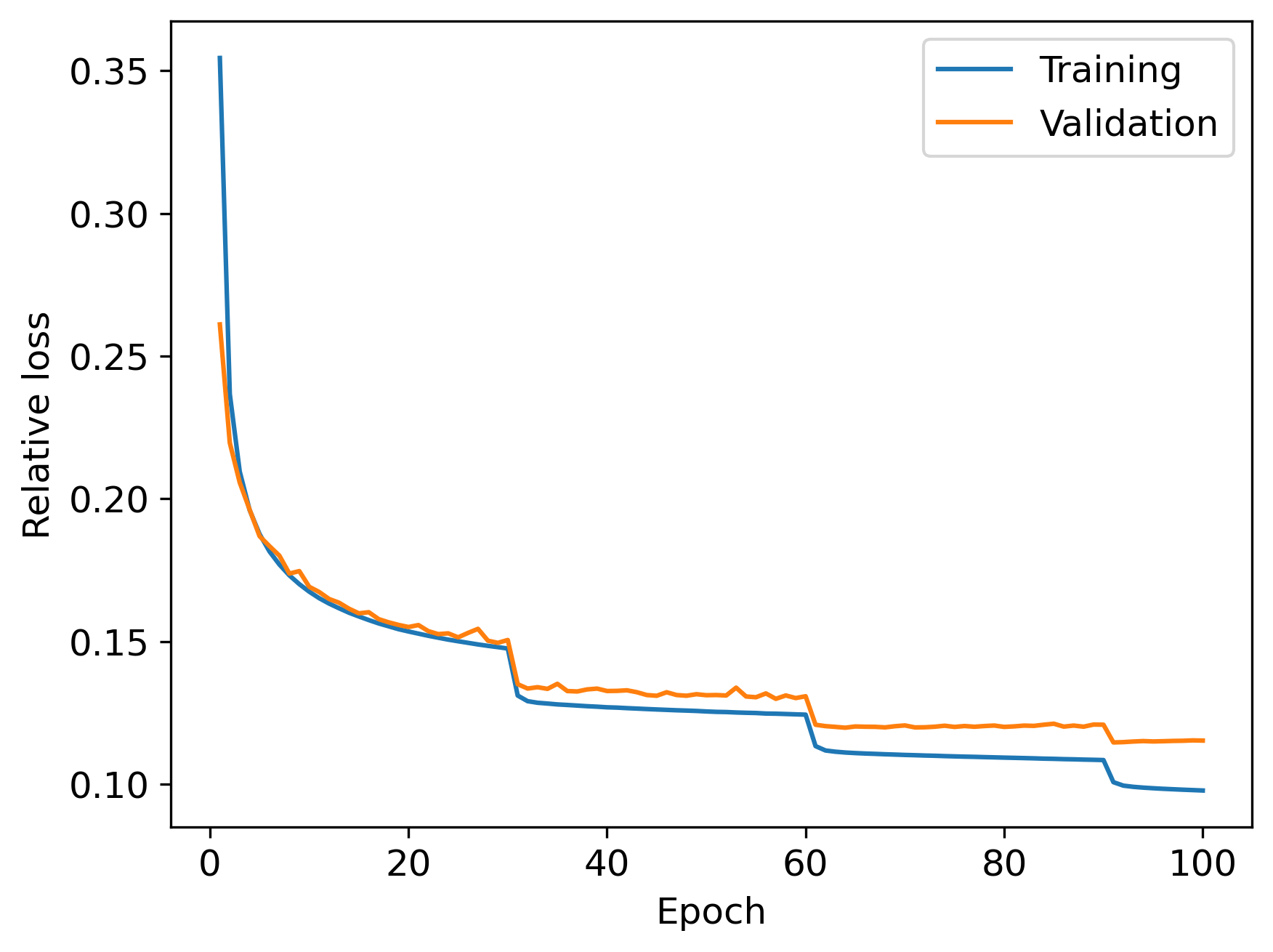} 
\caption{Loss curves during the 2D U-NO training for 100 epochs. The loss function is defined as the relative L1 norm with a weight of 0.9 plus the relative L2 norm with a weight of 0.1. The batch size is 32. An Adam optimizer is employed with a learning rate of 0.001 and a learning rate scheduler that decays the learning rate by half every 30 epochs.}
\label{loss2D}
\end{figure}

The 2D U-NO has 65,652,068 trainable parameters taking up 0.24 GB. The model architecture, depicted in the Methods section, is determined through trial and error based on the validation performance. The other hyperparameters are also optimized with the validation set. To strike a balance between accuracy and robustness, we adopt a loss function comprising the relative L1 norm with a weight of 0.9 plus the relative L2 norm with a weight of 0.1. A batch size of 32 is used for neural operator training. We employ an Adam (adaptive moment estimation) \citep{Kingma} optimizer with a learning rate of 0.001 and a learning rate scheduler that decays the learning rate by half every 30 epochs. The machine learning model is trained for 100 epochs until the validation set converges well, as shown in Fig. \ref{loss2D}. Each epoch takes around 15 minutes using eight NVIDIA RTX A6000 GPUs, each with an average memory usage of 1.3 GB. Note that the model can be trained with just a single GPU, but here we want to take advantage of PyTorch’s parallel computing capability, especially for the many frequencies that can be handled independently. Once the U-NO is trained, one run of the 2D elastic wave modeling with new velocity parameters and source locations for all the frequencies of interest only takes 0.03 s using one GPU with a memory usage of 0.7 GB, achieving a 53-fold acceleration in comparison to the SEM solver.

\begin{figure}
\centering
\includegraphics[width=0.7\textwidth]{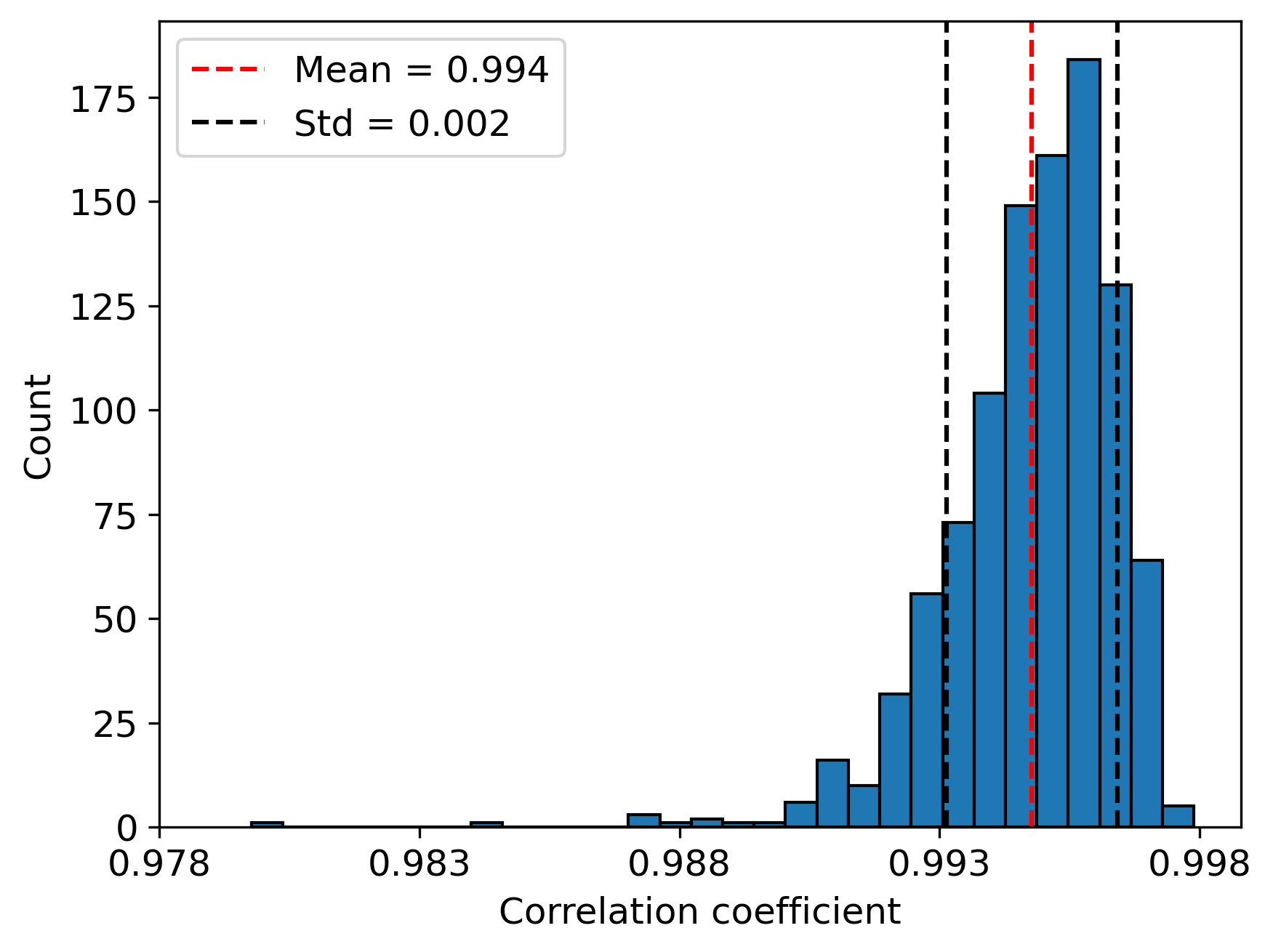}
\caption{Distribution of correlation coefficients between the 2D U-NO predictions and ground truth in the time domain for the test set. The red and black dashed lines indicate the mean and standard deviation of the histograms.}
\label{CC2D}
\end{figure}

Fig. \ref{loss2D} shows that the U-NO is able to fit the training data well and has an impressive predictive performance on the validation set after hyperparameter optimization. However, the generalization ability of the trained U-NO should be evaluated on a separate test set excluded from either model parameter training or hyperparameter optimization. The test set consists of 1000 simulations in the time domain, which after the Fourier transform becomes 12000 input-output pairs in the frequency domain. The trained U-NO achieves a relative test loss of 0.116 in the frequency domain. Because the relative loss tends to be dominated by small amplitudes, frequencies with less energy can be overweighted. For a fairer evaluation, the predicted results are also assessed in the time domain through cross-correlation with the ground truth simulated by the SEM. The correlation coefficient is more robust to small errors in Fourier phase. Fig. \ref{CC2D} shows the distribution of correlation coefficients between the 2D U-NO predictions and ground truth in the time domain for the test set. The 2D predictive performance is excellent with a mean correlation coefficient of 0.994, and all of the 1000 simulations have correlation coefficients over 0.98.

\begin{figure*}
    \centering
    \includegraphics[width=1\textwidth]{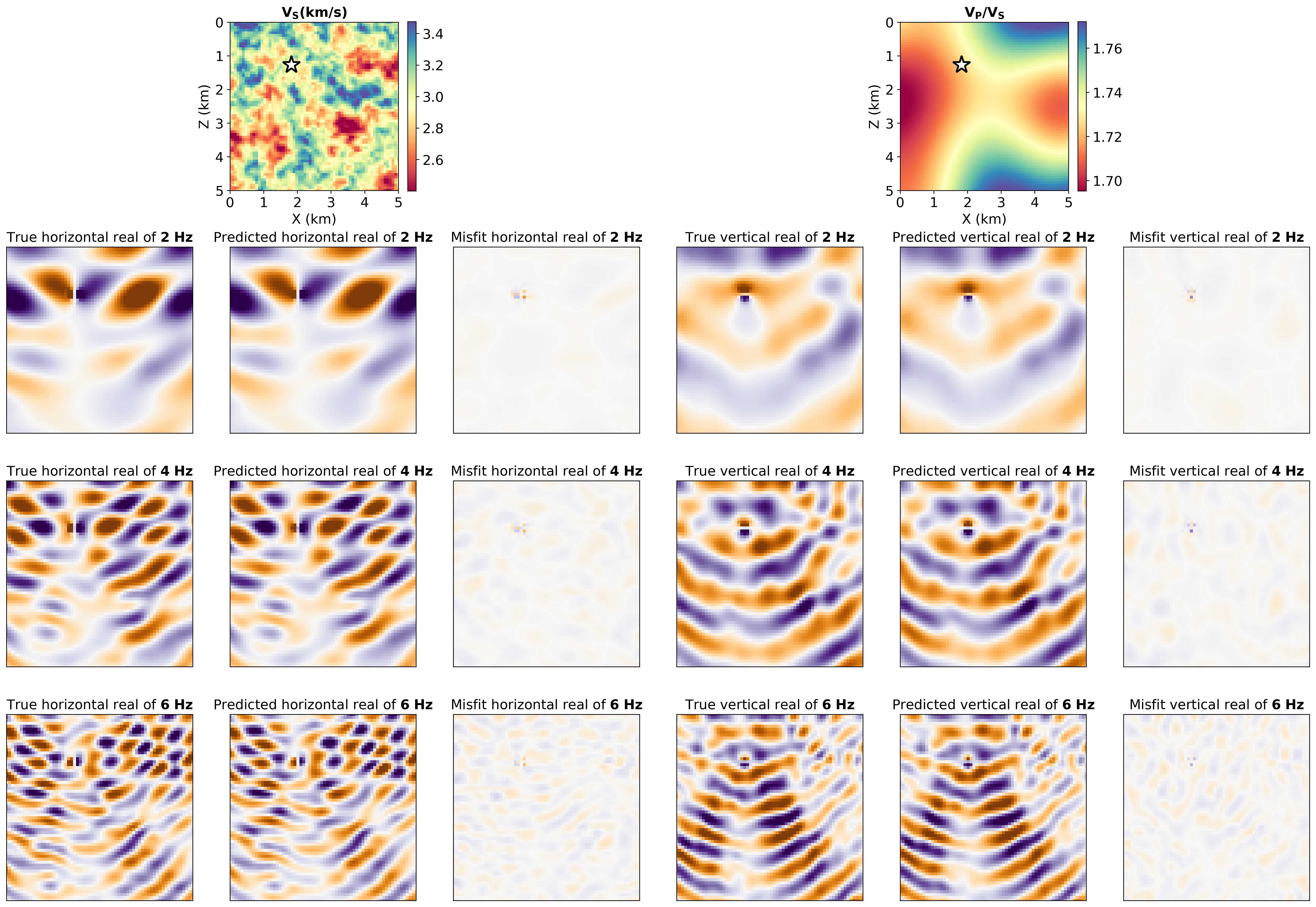}
    \caption{2D elastic wave modeling with U-NO evaluated in the frequency domain for an instance from the test set. The first row displays the $V_P$ and $V_P/V_S$ models with the source location marked with a white star. The second, third, and fourth rows display the predicted results for the real parts of displacement fields of 2 Hz, 4 Hz, and 6 Hz, respectively. The relative loss of the U-NO prediction is 0.060 for 2 Hz, 0.077 for 4 Hz, and 0.126 for 6 Hz.}
    \label{f2Dreal}
\end{figure*}

\begin{figure*}
    \centering
    \includegraphics[width=1\textwidth]{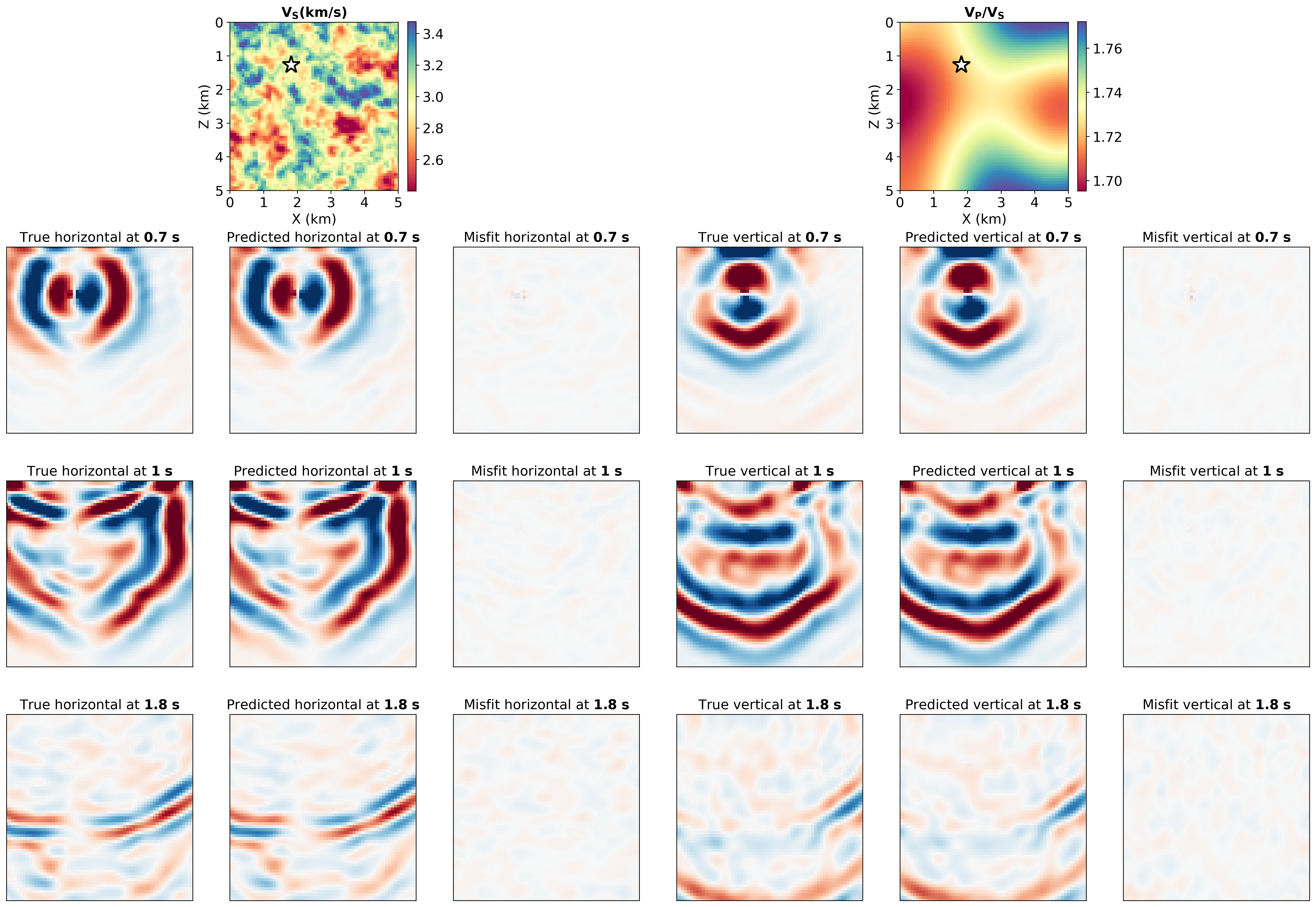}
    \caption{2D elastic wave modeling with U-NO evaluated in the time domain for an instance from the test set. The first row displays the $V_P$ and $V_P/V_S$ models with the source location marked with a white star. The second, third, and fourth rows display the predicted results for displacement fields at 0.7 s, 1 s, and 1.8 s, respectively. The cross-correlation coefficient between the U-NO and SEM simulations is 0.997.}
    \label{t2D}
\end{figure*}

We can also evaluate the performance of U-NO at a frequency level. Fig. \ref{f2Dreal} shows 2D Helmholtz solutions with U-NO for displacement fields of 2 Hz, 4 Hz, and 6 Hz for an instance from the test set. The relative losses for these cases are 0.060, 0.077, and 0.126, respectively. We only show the real parts of the wavefields here for brevity, and the imaginary parts are given in supplementary materials (Fig. \ref{f2Dimag}). The U-NO predictive accuracy depends on the given frequency and intuitively, higher frequencies are harder to learn. This will be discussed in detail in the following section. For the time-domain evaluation, this example demonstrates the excellent predictive capability of neural operators for seismic wave modeling with a correlation coefficient of 0.997 (Fig. \ref{t2D}). The success in 2D serves as a stepping-stone towards the exciting application of neural operators for 3D elastic wave modeling.

\subsection{3D elastic wave modeling with U-NO}
\begin{figure}
\centering
\includegraphics[width=0.7\textwidth]{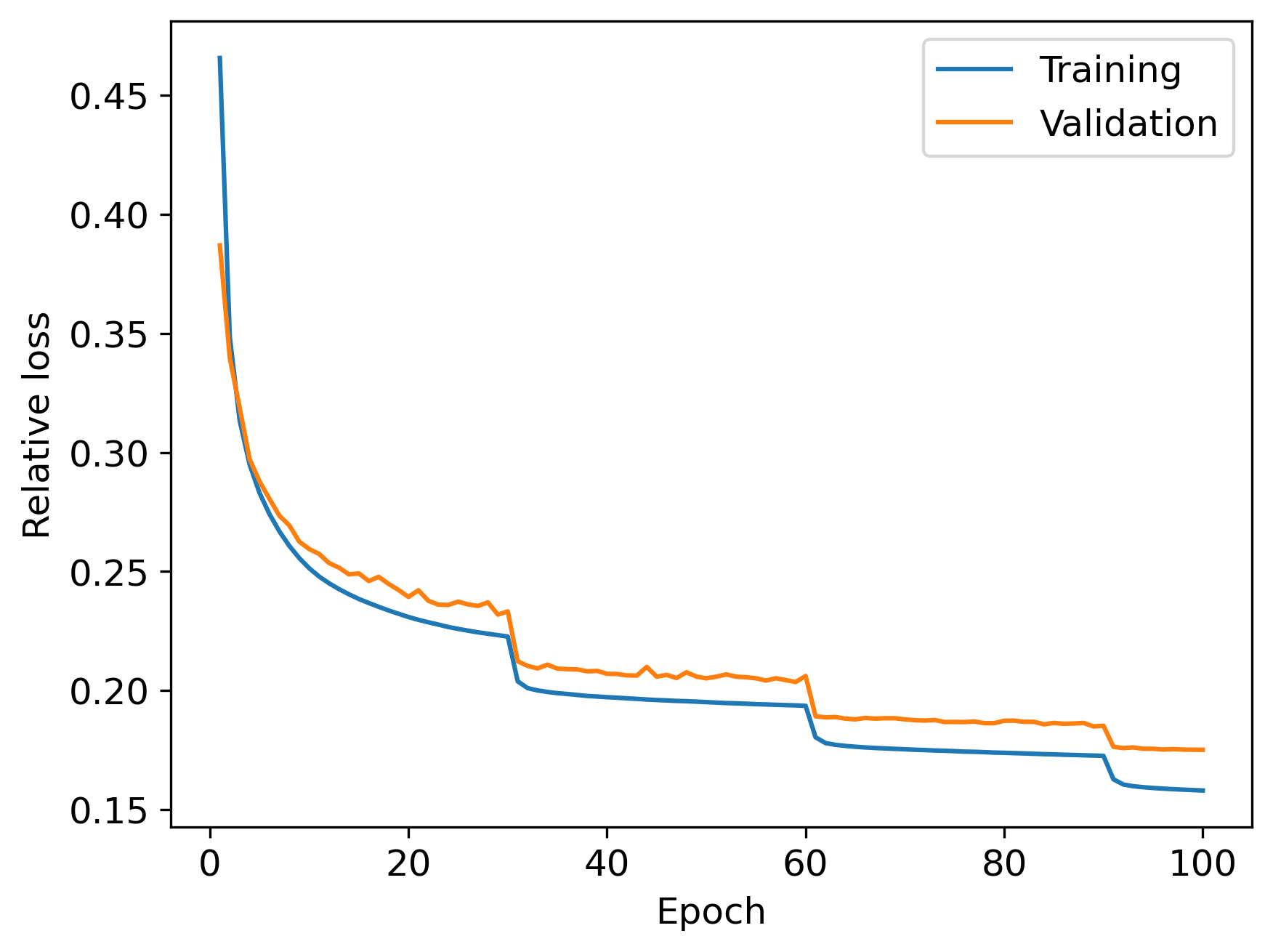}
\caption{Loss curves during the 3D U-NO training for 100 epochs. The loss function is defined as the relative L1 norm with a weight of 0.9 plus the relative L2 norm with a weight of 0.1. The batch size is 32. An Adam optimizer is employed with a learning rate of 0.001 and a learning rate scheduler that decays the learning rate by half every 30 epochs.}
\label{loss3D}
\end{figure}

Through hyperparameter optimization, the 3D model has 1,453,400,214 trainable parameters taking up 5.4 GB. Each epoch of model training takes about 3.7 hours using eight NVIDIA RTX A6000 GPUs, using an average of 16 GB memory during the training process after including the data and other overhead. The 3D loss curves during training are given in Fig. \ref{loss3D}, which demonstrate similar training behavior to the 2D model. It is necessary to mention that the learning rate scheduler has an appreciable influence on the training performance and can be considered a critical component. Although the training stage is time-consuming, it is only done once. With U-NO, one 3D forward simulation for all the frequencies of interest (explained in the Data section) takes 0.3 s with one GPU, while using 20 GB of GPU memory. This is about 100 times faster than the classic SEM and allows for more computationally efficient 3D FWI. The relatively heavy GPU memory usage arises from the fact that we evaluate the Helmholtz solutions for the full set of frequencies desired in parallel, which contrasts with the time stepping scheme of the SEM that only processes one time step at a time. One can conserve memory by solving for the Helmholtz solutions sequentially, but this comes at the expense of increased processing time.

\begin{figure*}
    \centering
    \includegraphics[width=1\textwidth]{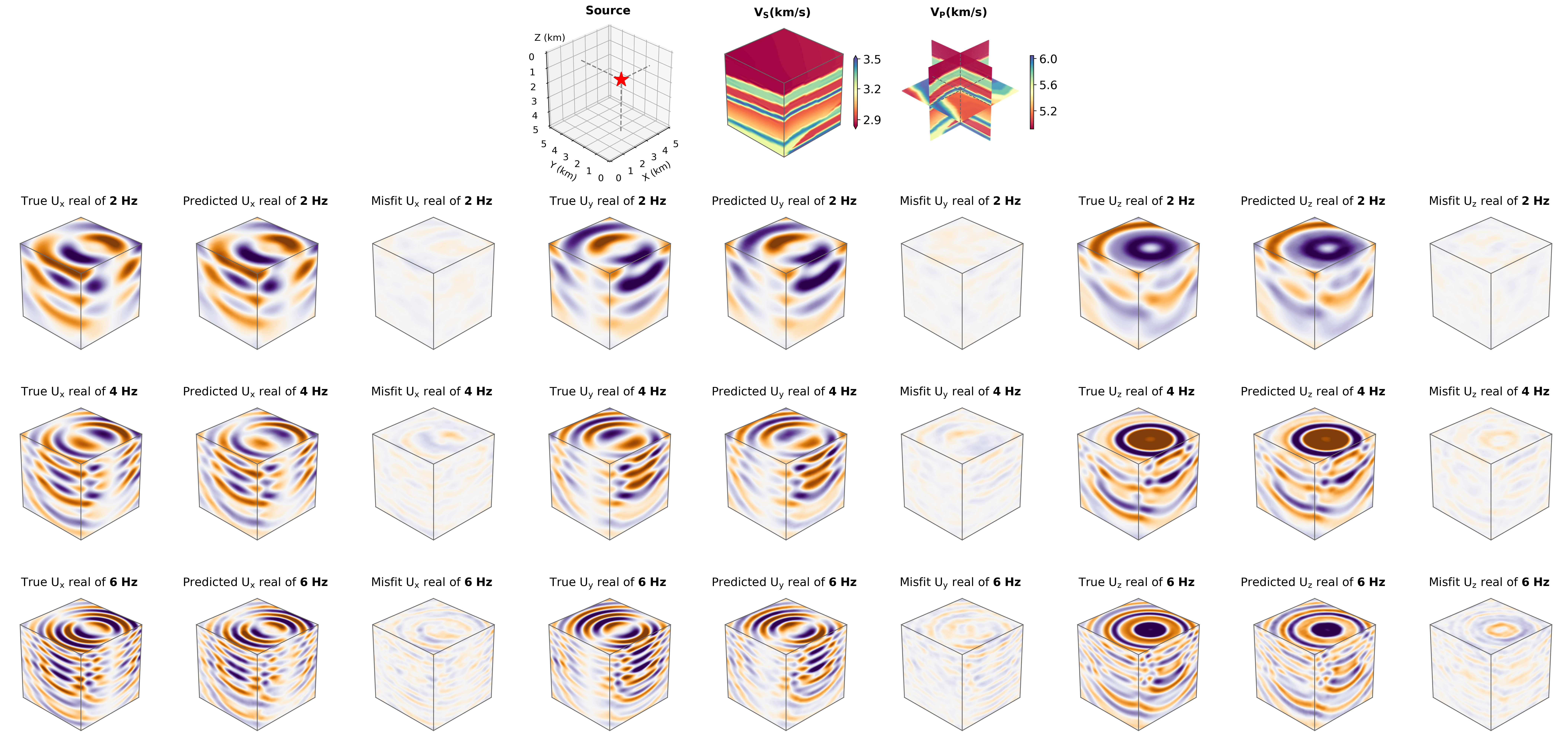}
    \caption{3D elastic wave modeling with U-NO evaluated in the frequency domain for an instance of random subpanels from the overthrust model. The first row displays the $V_P$ and $V_P/V_S$ models with the source location marked with a red star. The second, third, and fourth rows display the predicted results for the real parts of displacement fields of 2 Hz, 4 Hz, and 6 Hz, respectively. The relative loss of the U-NO prediction is 0.110 for 2 Hz, 0.154 for 4 Hz, and 0.241 for 6 Hz.}
    \label{f3Doverthrustreal}
\end{figure*}

\begin{figure*}
    \centering
    \includegraphics[width=1\textwidth]{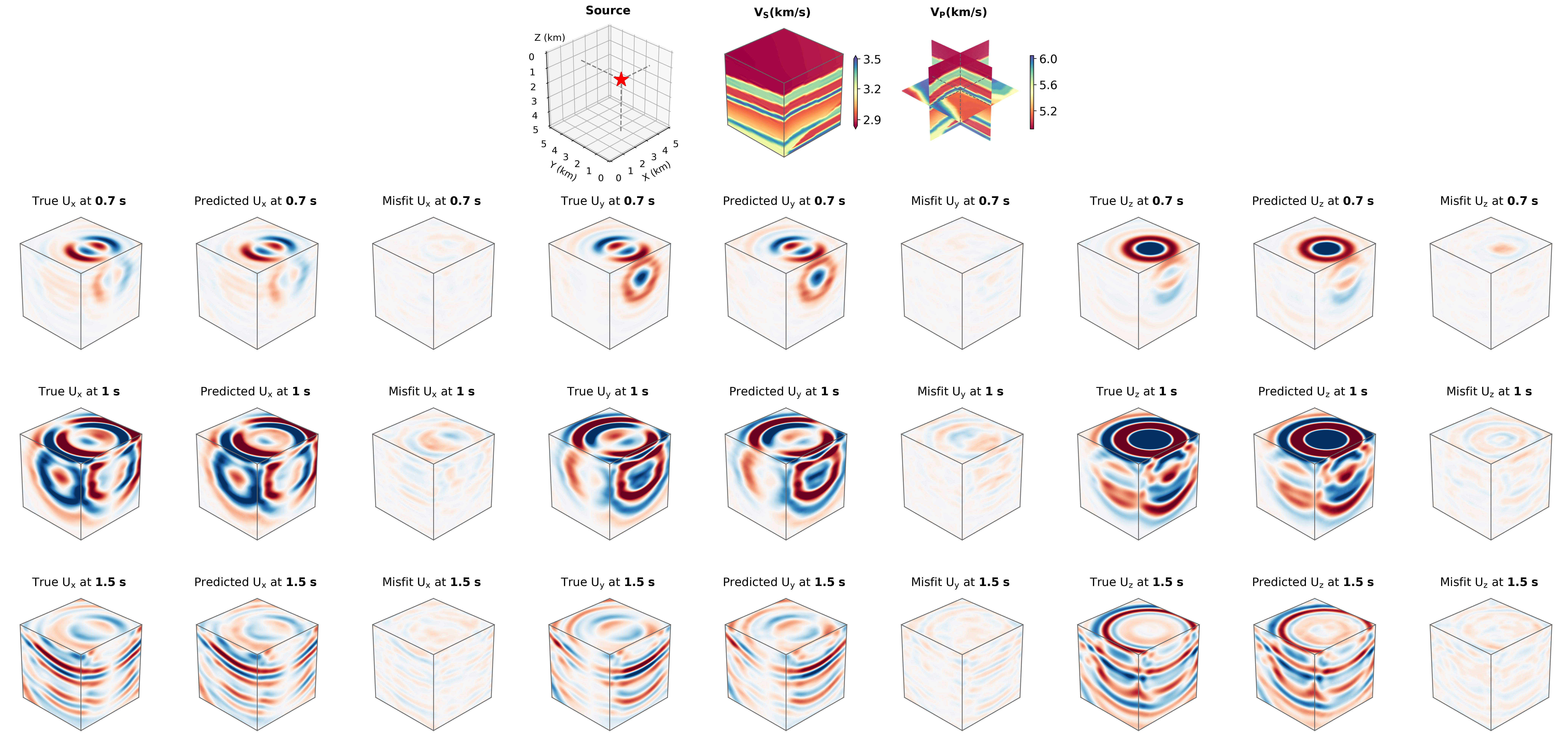}
    \caption{3D elastic wave modeling with U-NO evaluated in the time domain for an instance of random subpanels from the overthrust model. The first row displays the $V_P$ and $V_P/V_S$ models with the source location marked with a red star. The second, third, and fourth rows display the predicted results for displacement fields at 0.7 s, 1 s, and 1.5 s, respectively. The cross-correlation coefficient between the U-NO and SEM simulations is 0.991.}
    \label{t3Doverthrust}
\end{figure*}

\begin{figure*}
    \centering
    \subfloat[]{\includegraphics[width=0.4\textwidth]{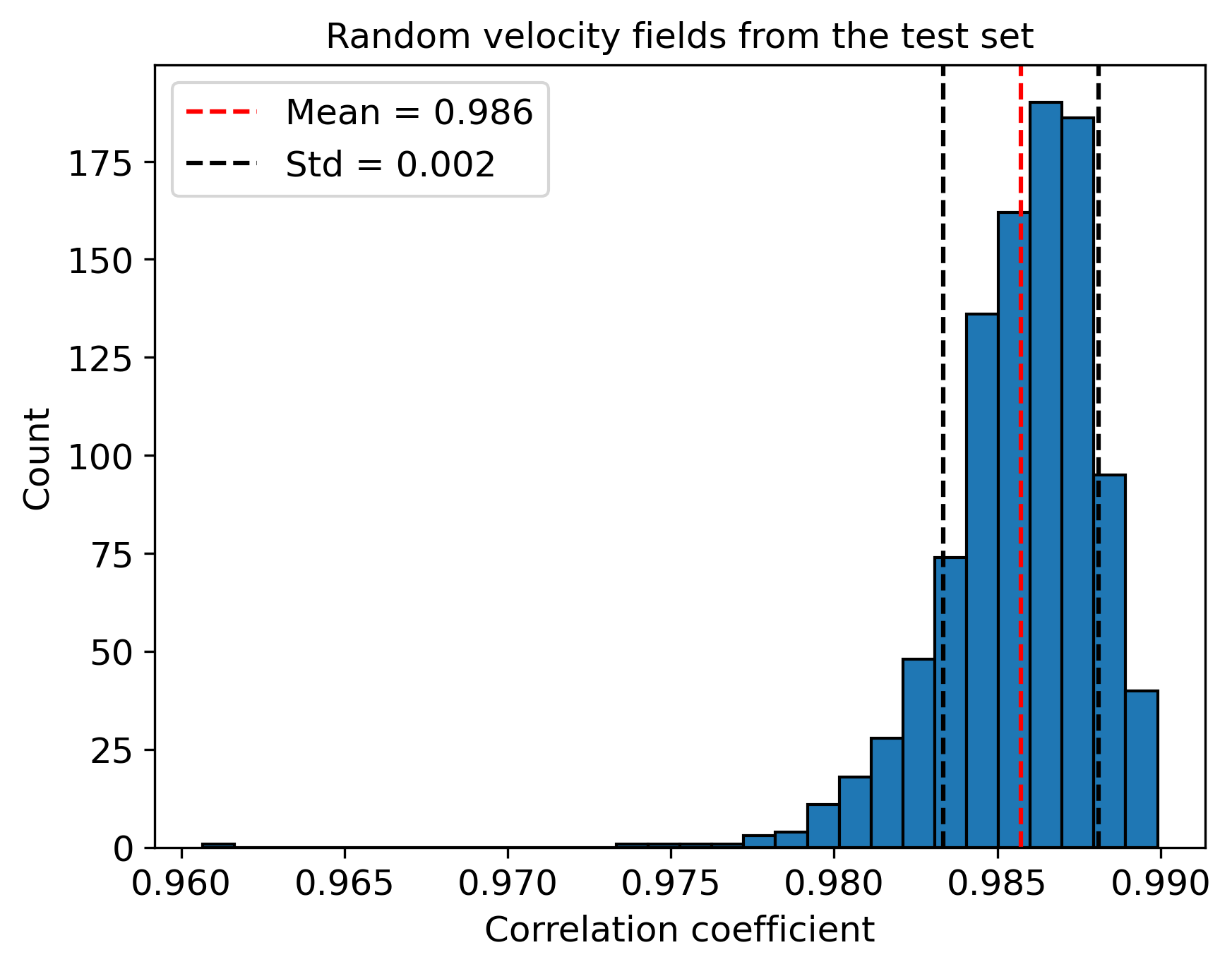}}
    \subfloat[]{\includegraphics[width=0.4\textwidth]{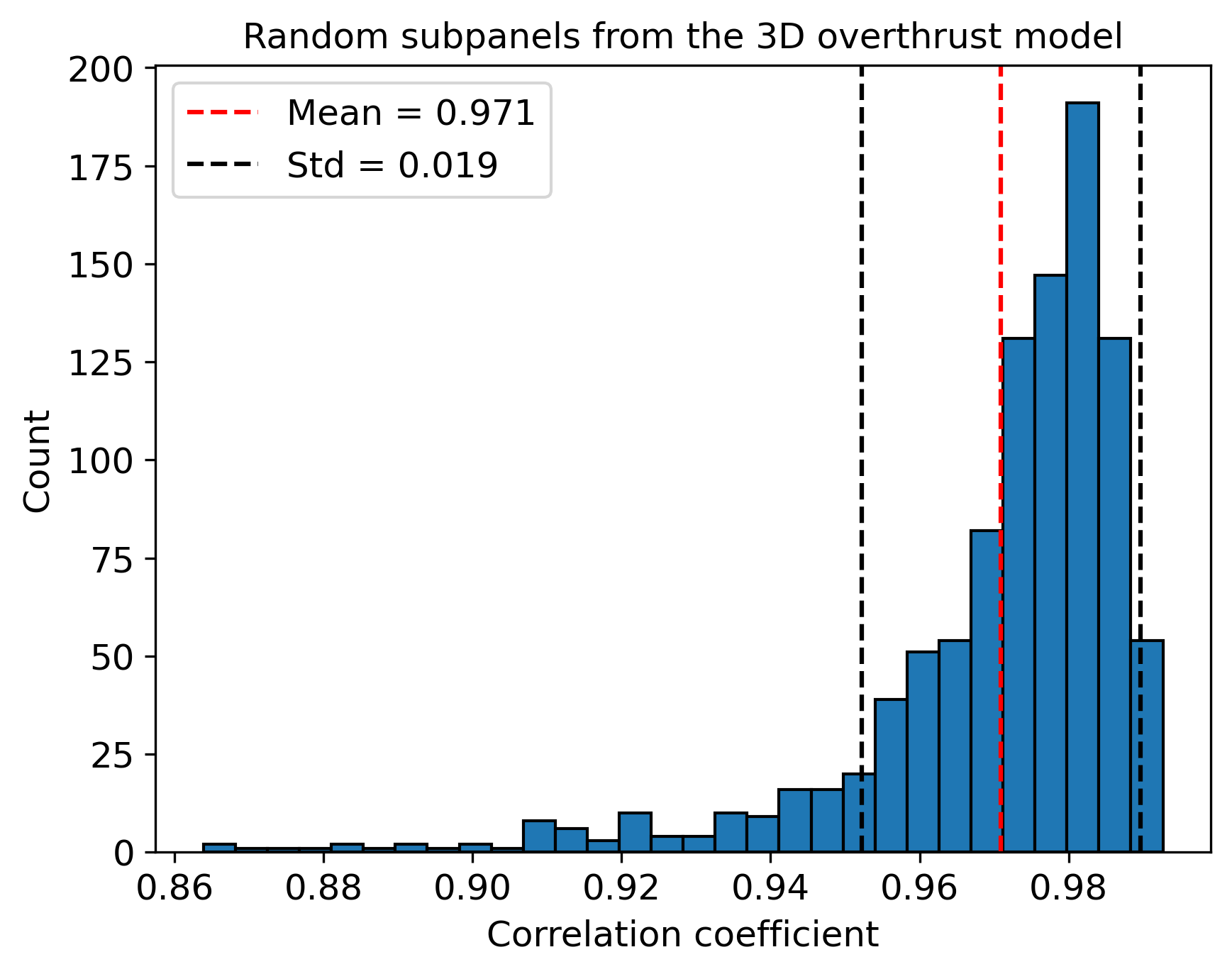}}
    \caption{Distribution of correlation coefficients between the 3D U-NO predictions and ground truth in the time domain for (a) random velocity fields from the test set and (b) random subpanels from the 3D overthrust model. The red and black dashed lines indicate the mean and standard deviation of the histograms.}
    \label{CC3D}
\end{figure*}

In addition to the random velocity fields, we incorporate an overthrust model commonly used in exploration studies \citep{Aminzadeh} into the generalization test for the 3D U-NO. 1000 random subpanels are extracted from the 3D overthrust model and a perturbation range of 30\% is imposed on an average $V_S$ of 3 km/s. The $V_P$ models are calculated from the $V_S$ models multiplied by 1.732. The density is set to 2.7 g/cm³ and the source is randomly placed in the computational domain for both test sets. The overall performance of the 3D U-NO on the test set of random velocity fields is 0.176 measured by the relative loss in the frequency domain, and 0.238 on the test set of overthrust models. Fig. \ref{f3Doverthrustreal} shows the 3D predicted results for the real parts of 2-, 4-, and 6-Hz wavefields for a subpanel from the overthrust model. The imaginary parts are given in supplementary materials (Fig. \ref{f3Doverthrustimag}). Fig. \ref{t3Doverthrust} illustrates the reconstructed wavefields for several snapshots in time, with the corresponding ground truth. It is hard to distinguish between the U-NO-predicted waveforms and SEM-simulated waveforms by eye. Importantly, the neural operator learns the diffraction effects at the sharp discontinuities. Figs \ref{f3Dreal}, \ref{f3Dimag} and \ref{t3D} in supplementary materials show the predicted results for an instance of random velocity fields in the frequency and time domains, respectively. For the 1000 simulations in each test data set, the 3D U-NO achieves a mean correlation coefficient of 0.986 for random velocity fields and 0.971 for random subpanels from the overthrust model, as shown in Fig. \ref{CC3D}. The higher standard deviation observed in the overthrust case could be in part attributed to the model’s complex heterogeneities. In general, the evaluation metrics demonstrate the promising generalization power of U-NO for real-world applications.

\subsection{Full-waveform inversion with automatic differentiation}
\begin{figure}
\centering
\includegraphics[width=0.8\textwidth]{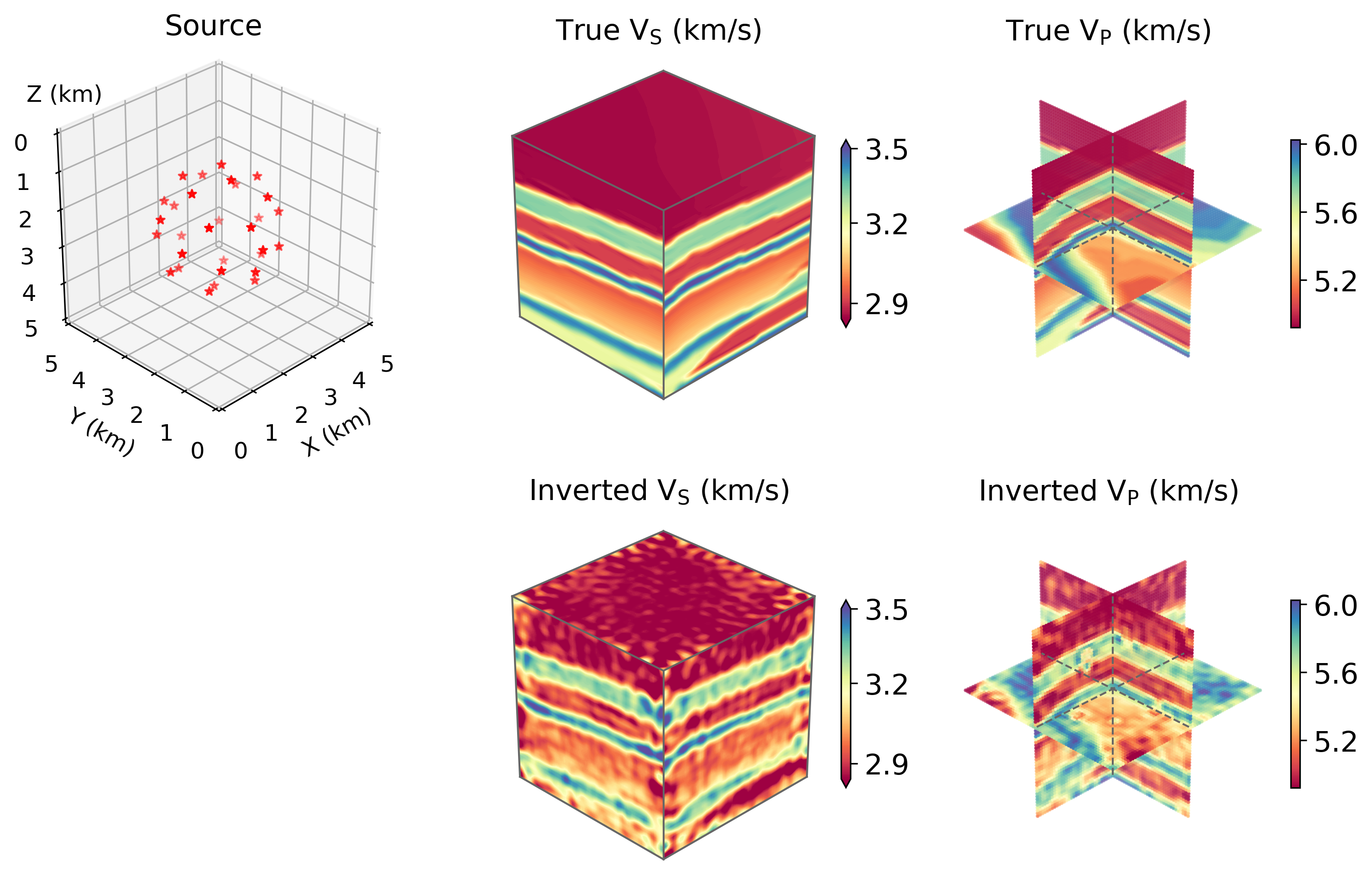} 
\caption{U-NO based full-waveform inversion for a random subpanel from the overthrust model with automatic differentiation. 30 sources are evenly distributed on the surface of a sphere with a radius of 1.5 km (red stars) and receivers are configured at every grid point of the 64×64×64 mesh. The true $V_S$ and $V_P$ models are plotted in the first row. The inverted velocity models with Laplacian smoothness regularization are plotted in the second row. The relative L2 norm of the misfit between the true and inverted $V_S$ and $V_P$ is 0.03.}
\label{fwifull}
\end{figure}

Full-waveform inversion (FWI) is a prime beneficiary of the accelerated seismic wave simulations, as it uses the complete information of recorded waveforms to infer subsurface physical parameters sampled by seismic waves. The adjoint-state method is conventionally used for FWI, which requires derivation of the gradient of the objective function with respect to each parameter of interest \citep{Plessix}. A counterpart to the adjoint-state method has been discovered in the deep learning community, known as automatic differentiation (AD). It has been shown that these two methods are mathematically equivalent \citep{LeCun,Zhu}. The advantage of AD over the adjoint-state method is that it automatically computes gradients for any desired parameter based on the computational graph using the chain rule, integrating a wide range of inverse problems in a unified framework. Neural operators are inherently designed to be compatible with automatic differentiation. In performing FWI, we freeze the U-NO model parameters and instead treat the velocity parameters as the objects of training. The objective function is the same as used for training the U-NO model, except we add a regularization term that encourages model smoothness using the Laplacian operator. We take the overthrust velocity models used in Fig. \ref{t3Doverthrust} for the FWI experiment. The ``observations'' are wavefields simulated with SALVUS for 30 events evenly distributed on the surface of a sphere with a radius of 1.5 km (Fig. \ref{fwifull}). Receivers are placed at each grid point of the 64×64×64 mesh. The $V_P$ and $V_S$ models are simultaneously updated, starting from homogeneous initial values of 5 km/s and 3 km/s, respectively. After conducting 10 epochs of training using an Adam optimizer with a learning rate of 0.03 and a batch size of 16, we achieve $V_P$ and $V_S$ models with relative L2 misfit of 0.03, as depicted in Fig. \ref{fwifull}. 

\begin{figure}
\centering
\includegraphics[width=0.8\textwidth]{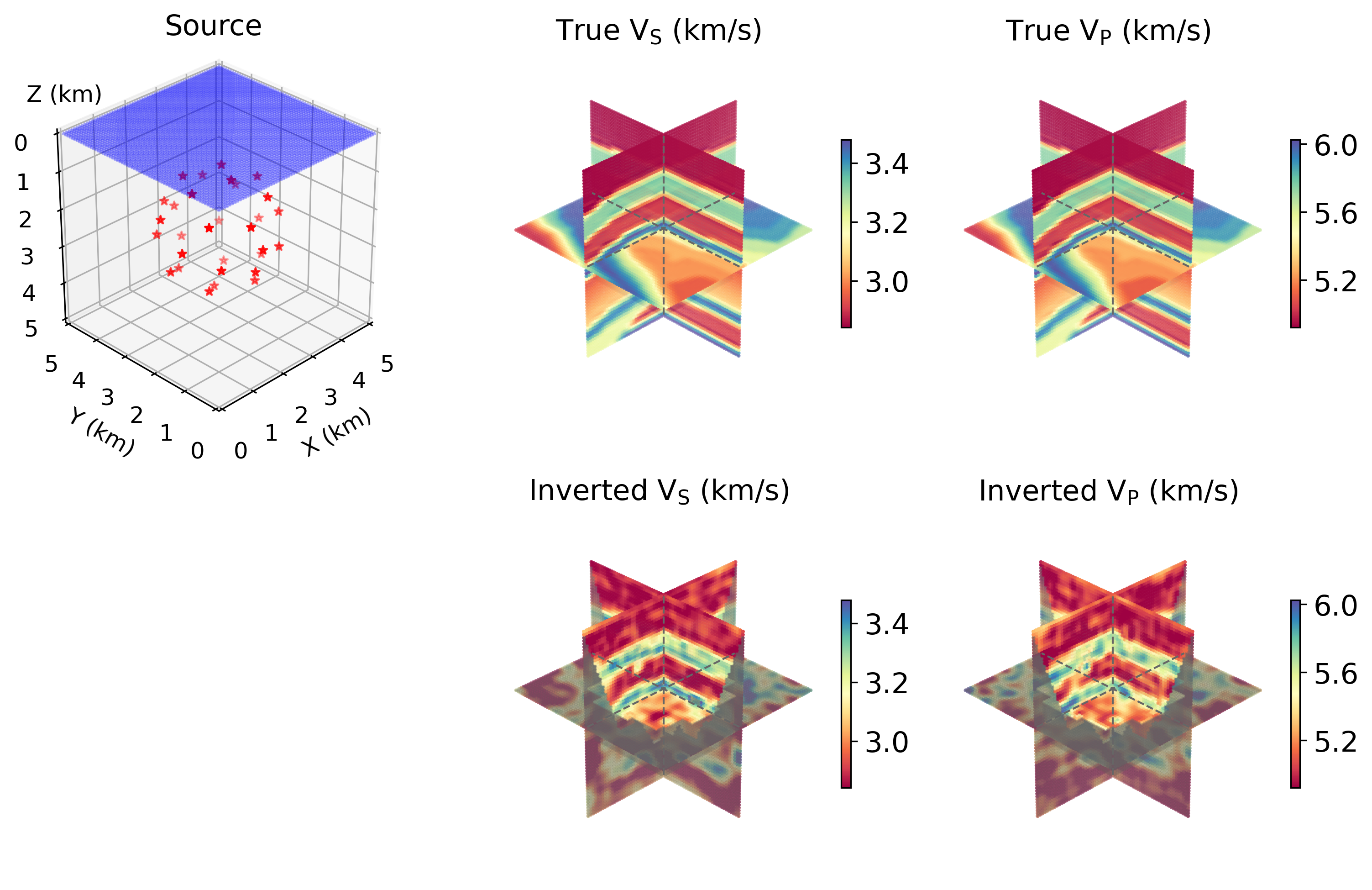}
\caption{U-NO based full-waveform inversion for a random subpanel from the overthrust model with automatic differentiation. 30 sources are evenly distributed on the surface of a sphere with a radius of 1.5 km (red stars) and 64×64 receivers are configured on the surface (blue area). The true $V_S$ and $V_P$ models are plotted in the first row. The inverted velocity models with Laplacian smoothness regularization are plotted in the second row, where parts without ray path coverage are masked by gray shadows. The relative L2 misfit for the ray-covered parts is 0.03.}
\label{fwisurf}
\end{figure}

This FWI experiment validates the effectiveness of automatic differentiation as a substitute for the adjoint-state method, and further showcases the accuracy of 3D elastic wave modeling with U-NO. We also conduct a similar experiment with receivers only on the surface (Fig. \ref{fwisurf}). In plotting the results, we mask out the regions with poor ray coverage as they are difficult to infer. The regions with sufficient ray density still obtain satisfactory inversion accuracy, with relative L2 misfit of 0.03. For one event and one tomographic iteration, the U-NO-based FWI takes 0.4 s (including the forward propagation) using one GPU, while the SEM with the adjoint method takes 140 s under the same settings. This translates to a speed-up of 350 times.

\subsection{Generalization to denser input/output meshes}
\begin{figure*}
\centering
\includegraphics[width=1\textwidth]{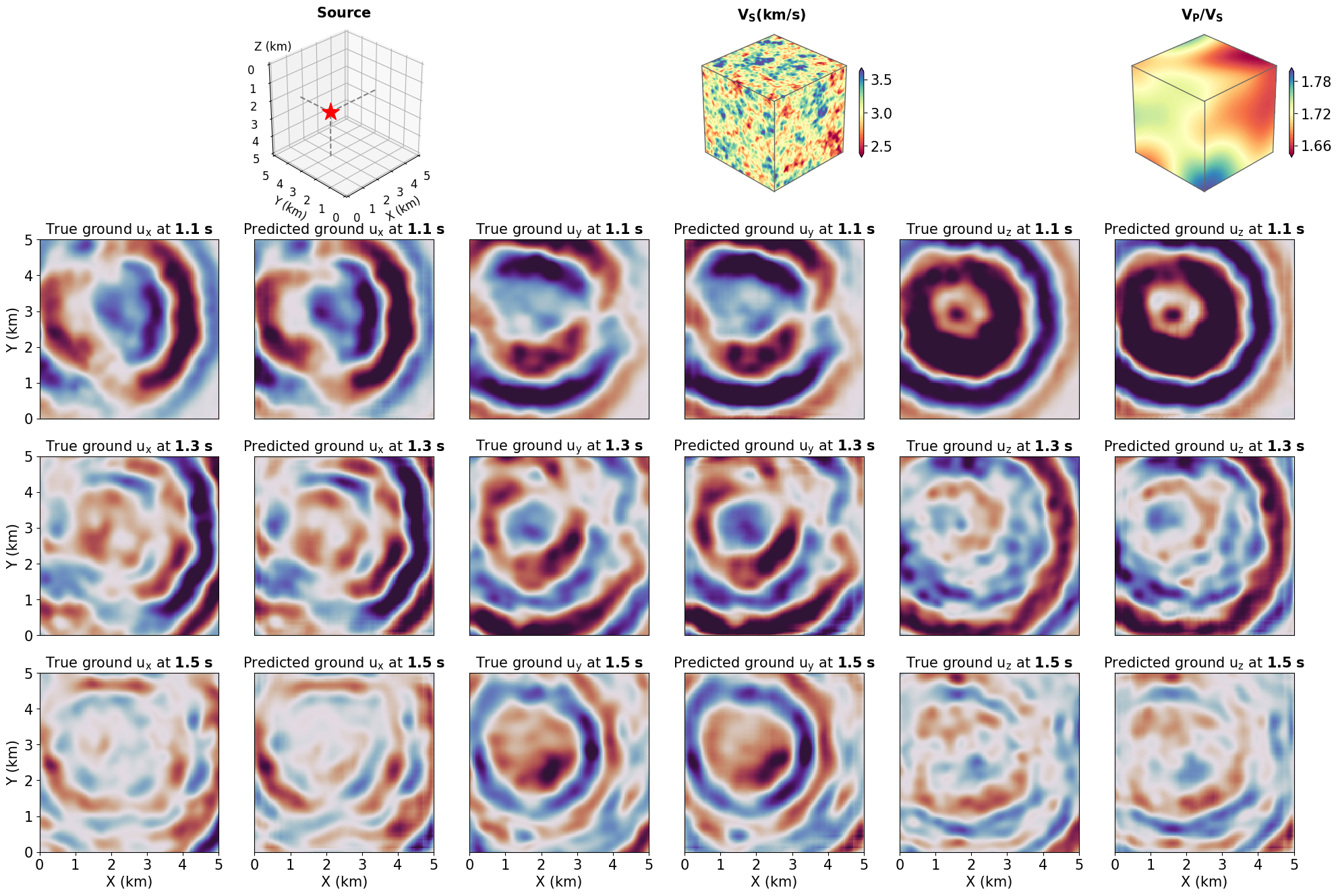}
\caption{3D elastic wave modeling on a 128×128×64 mesh with the U-NO trained on a 64×64×64 mesh. The first row displays the $V_P$ and $V_P/V_S$ models from random fields, with the source location marked with a red star. The second, third, and fourth rows display the true and predicted ground motions at 1.1 s, 1.3 s, and 1.5 s, respectively. The cross-correlation coefficient between the U-NO and SEM simulations for full wavefields of this example is 0.977.}
\label{super}
\end{figure*}

While classic neural networks are mesh-dependent, neural operators are discretization-invariant in physical space and can be evaluated on denser input and output meshes after training has concluded  \citep{Li2020a}. One of the most valuable aspects about this is that the training simulations do not need to be performed at the target resolution, which may be too computationally expensive. Here, we test this critical property of neural operators by applying the U-NO trained on a 64×64×64 mesh to a 128×128×64 mesh with finer spacing in the horizontal plane. Both the input velocity models and the output displacement wavefields are evaluated on denser grids. We conduct 100 experiments using random fields to get an average correlation coefficient of 0.971. Fig. \ref{super} shows one of the generalization experiments, with the ground motion plotted as an example. It can be seen that the U-NO can effectively generalize to denser discretization without additional training. For the 3D super-resolution test, the neural operator has a further speed-up of 144 times over the SEM solver in forward propagation.

\subsection{Predicting wavefields at only the free surface}
\begin{figure*}
    \centering
    \subfloat[]{\includegraphics[width=0.65\textwidth]{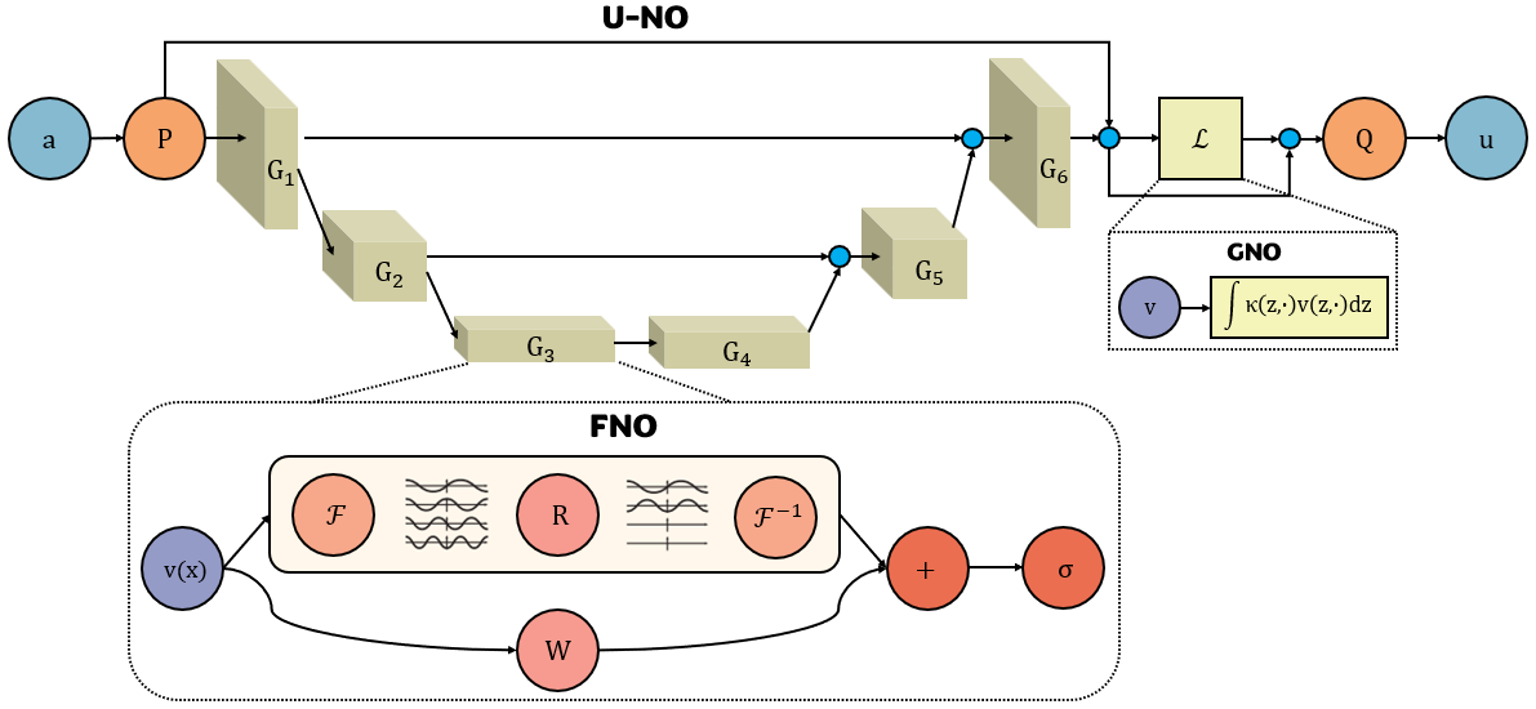}}
    \subfloat[]{\includegraphics[width=0.35\textwidth]{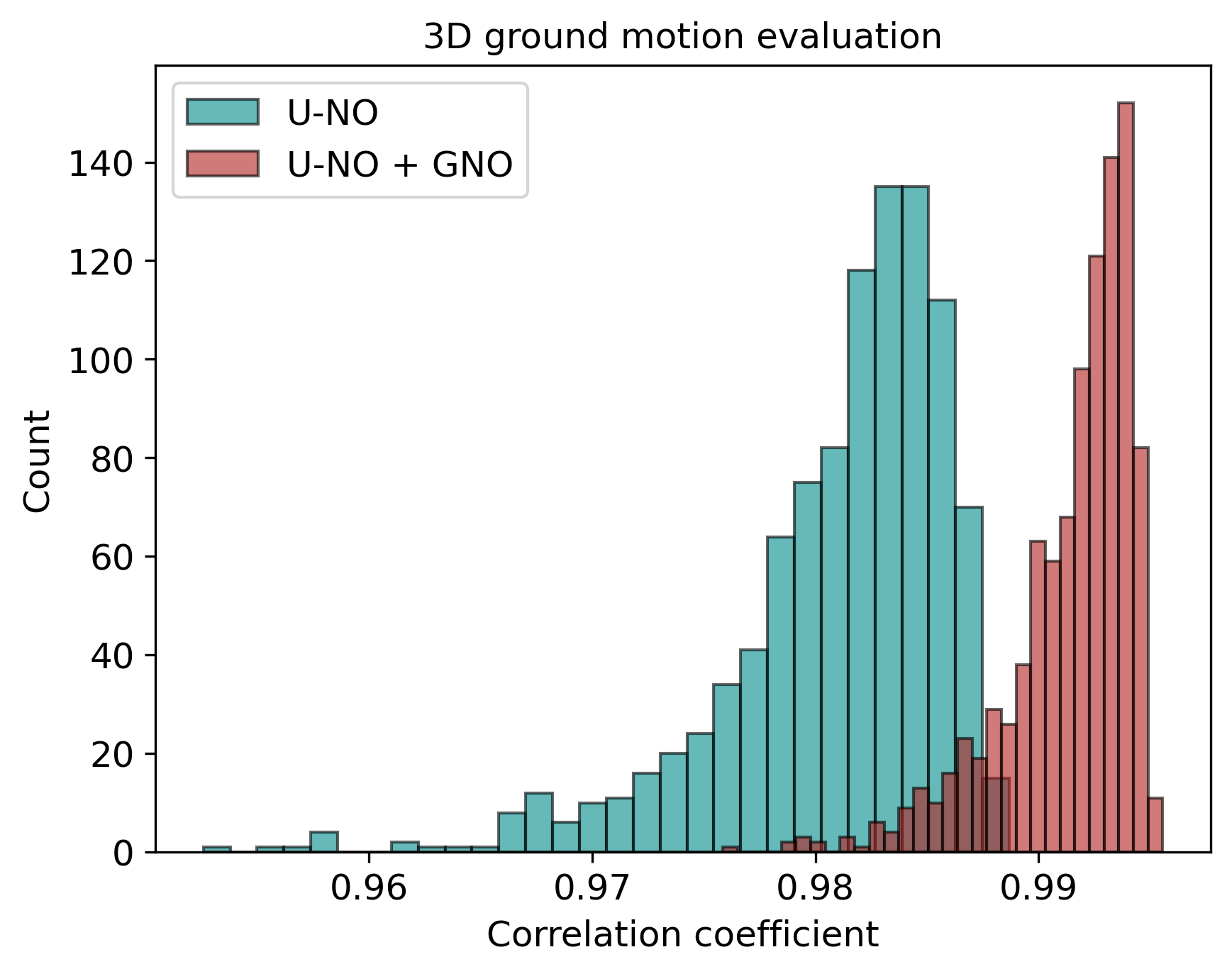}}
    \caption{(a) Model architecture for predicting wavefields at the free surface comprising the U-NO in Fig. \ref{uno} and a GNO that queries the ground motions.(b) Distribution of correlation coefficients between true and predicted 3D ground motions for 1000 random velocity models using the U-NO with (red) and without (cyan) a GNO layer.}
    \label{gno}
\end{figure*}

Thus far, we have employed the machine learning model to simulate wavefields over the full spatial domain. However, in most real-world cases we will only have access to waveform data at/near the surface. In the framework of supervised learning, it is technically feasible to directly map from elastic properties to surface ground motions directly at (irregular) receiver points. Under many circumstances, this could be more memory- and computationally-efficient. To achieve this, the model architecture needs to be slightly modified for generalizability. We add a GNO~\citep{Li2020b} layer to the top of the U-NO architecture to query the ground motions at the desired points in 3D space. GNO layers are evalulated using message passing graph neural networks \citep{Gilmer}. They incorporate positional information into edge features, allowing for the querying of solutions at arbitrary points on arbitrary discretizations, not limited to a planar free surface. Fig. \ref{gno}(a) gives the model architecture for predicting wavefields on the free surface, where the GNO employs a kernel function $\kappa _{l}$ parameterized as a three-layer neural network. $\kappa _{l}$ takes the positional information and the output of the U-NO as input. For the GNO, we let messages pass across the whole computational domain to the free surface through a linear integral. The remaining part of the model adheres to the U-NO architecture in Fig. \ref{uno}. We compare the predictive accuracy for 3-component surface wavefields using the plain U-NO and the GNO-embedded model. For the plain U-NO, the ground motions are output over the whole domain but evaluated strictly at the surface. We calculate the correlation coefficients for 1000 random velocity models unseen in training, as shown in Fig. \ref{gno}(b). Incorporating the GNO systematically improves surface wavefield predictions while also being more data-efficient. We re-emphasize that a GNO layer can query arbitrary points on arbitrary discretizations, not just the free surface. This is crucial for future applications having an irregular geometry of seismic stations at variable elevation.

\section{Discussion}
Excluding the dataset preparation, the most computationally expensive part of the proposed method is the model training process. Optimizing the model architecture and hyperparameters is quite time consuming and tedious. Rather than documenting every detail of this trial-and-error process, we summarize the key findings here: 
\begin{enumerate}
\item We find that skip connections play an important role in the predictive accuracy of U-NO. They also contribute to alleviating vanishing gradient issues. 
\item Expanding the size of the co-domain can improve the U-NO performance, but at the cost of higher memory usage. 
\item The number of Fourier modes in each layer is a sensitive hyperparameter and should be tuned carefully. In seismic wave propagation, the elastic media exhibit complex heterogeneities that amplify high-frequency components in the resulting wavefields. Therefore, we retain all the Fourier modes in respective dimensions. 
\item As with many deep learning studies, the batch size has an appreciable effect on the performance. 
\item Learning rate scheduling is very effective, which reduces the learning rate as training progresses. 
\item Lastly, in the specific setting described in this paper, neural operators are very data hungry. We use 27000 simulations (which make $\sim$300000 frequency-domain instances) to train each U-NO. The belief is that the increased data volume will enhance accuracy, but this also translates into additional time taken up by numerical simulations and U-NO training.
\end{enumerate}

The frequencies in this study are discrete values determined by the fast Fourier transform, as we use a time-domain solver to generate data. If a Helmholtz solver is available, the frequencies can be randomly picked from the real number line $\mathbb{R}$ for training and not restricted to a discrete set. Within a predefined frequency band $[f_{min},f_{max}]$, a well-trained neural operator can be queried at arbitrary frequencies for Helmholtz solutions, because it learns mappings between function spaces. This adapts well to the band-limited characteristics of real seismic data. An additional advantage of the proposed method over a classical time-domain solver emerges when extending the simulation duration. The time consumed by a conventional solver following the time-stepping scheme increases linearly with the simulation duration, theoretically. However, because extending duration translates to querying intermediate frequencies, the increased run time for the Helmholtz neural operator can be alleviated by parallelization. 

\begin{figure}
\centering
\includegraphics[width=0.7\textwidth]{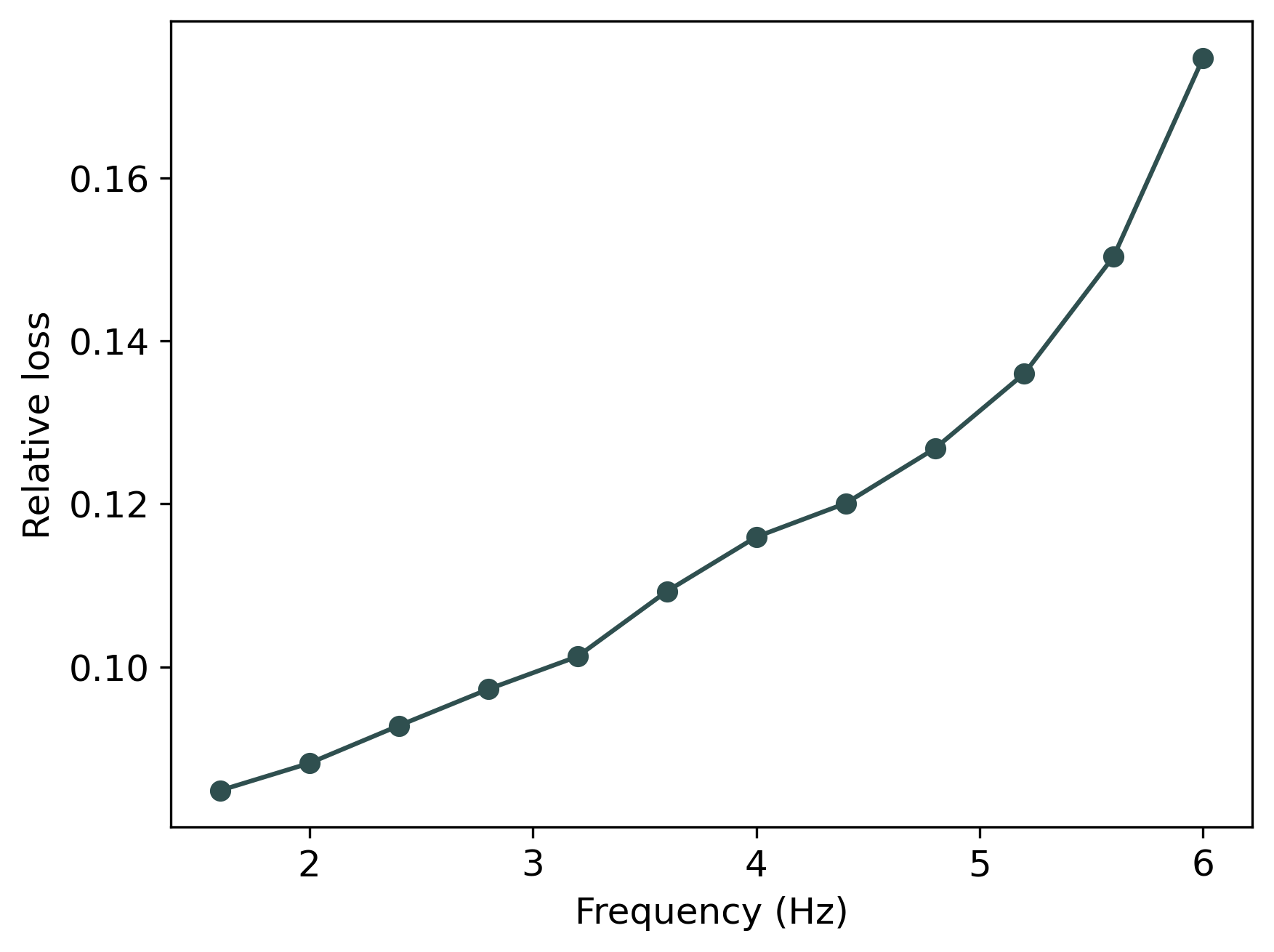}
\caption{Relative loss associated with frequencies for the 2D test set. The training set and 3D cases exhibit similar behavior.}
\label{lpf}
\end{figure}

In nearly all numerical methods, higher frequencies correspond to shorter wavelengths, necessitating finer gridding. Therefore, for any given same mesh, lower accuracy is expected for higher frequencies. This holds true for the machine learning method as well. We compute the relative loss with respect to each frequency of interest for the 2D test set, as shown in Fig. \ref{lpf}. It is evident that the loss monotonously goes up with increasing frequencies. Similar behavior is observed in the training set and 3D cases. While this arises naturally in part from the physics, it can also be partially attributed to the tendency of neural networks to generate smooth output to enhance generalizability \citep{Neal}. To reconcile the biased learning of neural operators towards low frequencies, future work can focus on exploring sampling methods, improved positional encoding, and weighted training \citep{Li2021a,Zhao}, among other potential approaches.

This study serves as a toolbox for applying neural operators to real data applications. In a realistic scenario, the physical domain could be larger, which places greater demands on computing efficiency. One way to reduce memory requirements and accelerate model training is to employ mixed precision training for neural operators \citep{White}. For now, our machine learning model still depends on a separate numerical solver, but a physics-informed version of neural operators \citep{Li2021b} has the potential to free the machine learning model from any external solver. In this study, we have given particular emphasis to the discretization invariance of neural operators and demonstrated it with an example of denser grids, but neural operators can also deal with irregular grids. Through a GNO layer, local predictions at arbitrary points (e.g., where there are seismic stations) can be queried and enhanced. For now, we have still kept the source as isotropic, but for earthquake scenarios, future work will involve parameterizing the source as different types of moment tensors.

Ultimately, we aim to provide a few models pretrained to high accuracy with very large datasets that can be shared as community resources. These models would be applicable to any regions on Earth as long as the study area fits within the constraints of the physical domain for which the model was trained. The training cost would therefore be a one time upfront expense, as future downstream users would not need to repeat this process.

\section{Conclusions}
We show that elastic wave modeling in 2D and 3D can be accelerated by two orders of magnitude with U-shaped Neural Operators. The prohibitive expense of memory-intensive models required in 3D is alleviated by modeling the seismic wave propagation in the frequency domain, due to the fact that multiple frequencies can be handled separately by parallel computing. We demonstrate the generalizability of the once-trained neural operator to variable velocity structures with random source locations, which is based on the theory that random fields can approximate most physical functions with arbitrary accuracy. The overall generalization accuracy is above 0.97 evaluated by cross-correlation coefficients, taking simulations with a spectral element method as ground truth. This sheds some light on the potential of the proposed method in real-world applications. We also show that neural operators can generalize to denser discretization without additional training. An additional GNO layer greatly improves the waveform predictions on the free surface. Moreover, we incorporate the trained model with automatic differentiation to facilitate rapid full-waveform inversion for velocity structures. The 3D inversion process can be accelerated by a factor of 350 in comparison to the SEM with the adjoint-state method, with accuracy measured by relative L2 misfit of 0.03. While being mathematically equivalent to the adjoint method, automatic differentiation can integrate a wide range of inverse problems in a unified framework without manual derivation.

\section*{Acknowledgements}
ZER thanks the David and Lucile Packard Foundation for supporting this study through a Packard Fellowship.

\bibliography{main.bib}
\bibliographystyle{myst}

\clearpage

\clearpage
\appendix
\input{supple}

\label{lastpage}

\end{document}

%% file: supple.tex
\allowdisplaybreaks

\setcounter{figure}{0} 
\setcounter{page}{1}

\renewcommand{\thepage}{S\arabic{page}} 
\renewcommand{\thefigure}{S\arabic{figure}}
\renewcommand{\thesection}{S\arabic{section}} 
\renewcommand{\theequation}{S.\arabic{equation}} 

\section*{Supplementary Materials}

\begin{figure*}[h]
    \centering
    \includegraphics[width=1\textwidth]{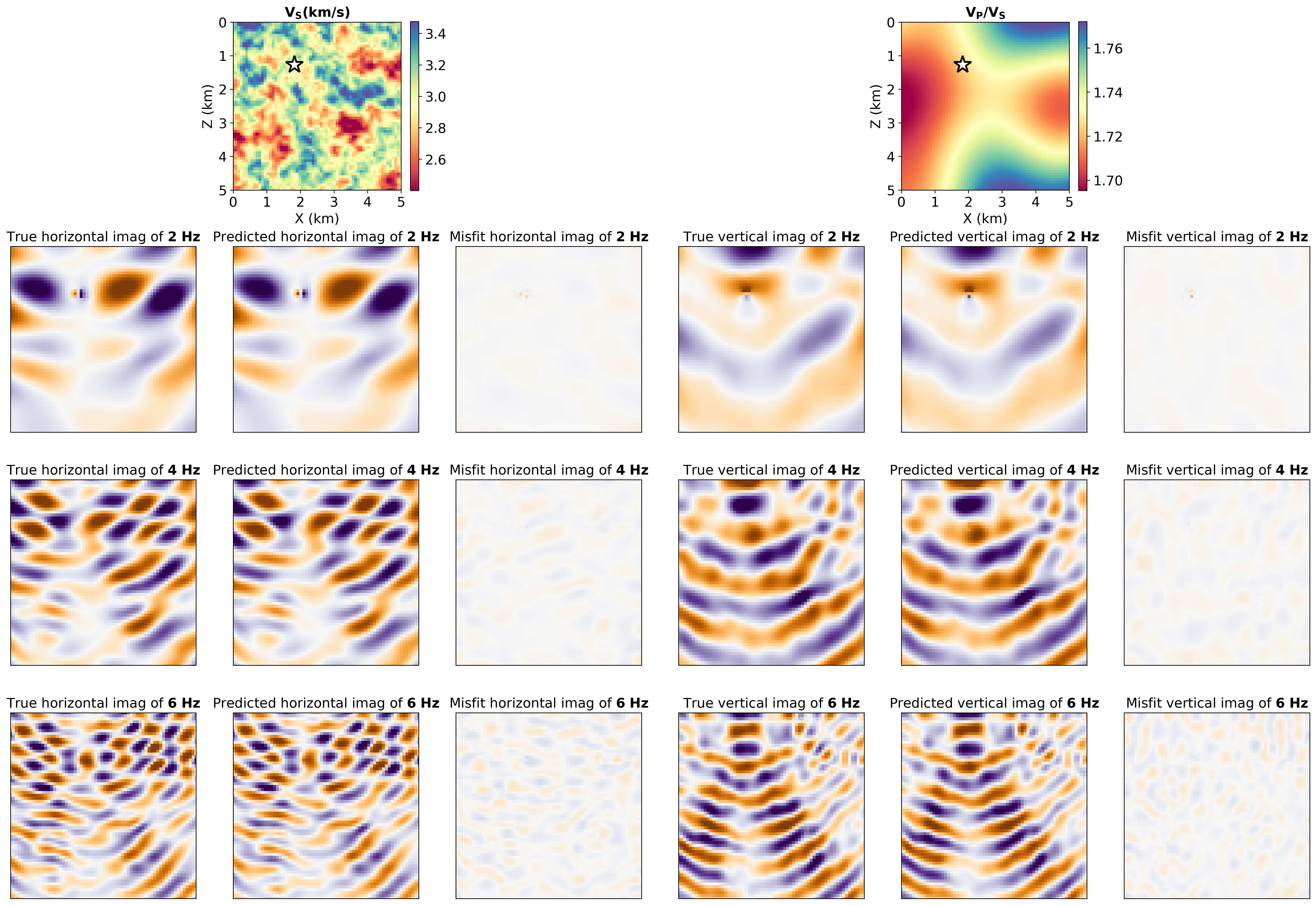}
    \caption{2D elastic wave modeling with U-NO evaluated in the frequency domain for an instance from the test set. The first row displays the $V_P$ and $V_P/V_S$ models with the source location marked with a white star. The second, third, and fourth rows display the predicted results for the imaginary parts of displacement fields of 2 Hz, 4 Hz, and 6 Hz, respectively. The relative loss of the U-NO prediction is 0.060 for 2 Hz, 0.077 for 4 Hz, and 0.126 for 6 Hz.}
    \label{f2Dimag}
\end{figure*}

\begin{figure*}
    \centering
    \includegraphics[width=1\textwidth]{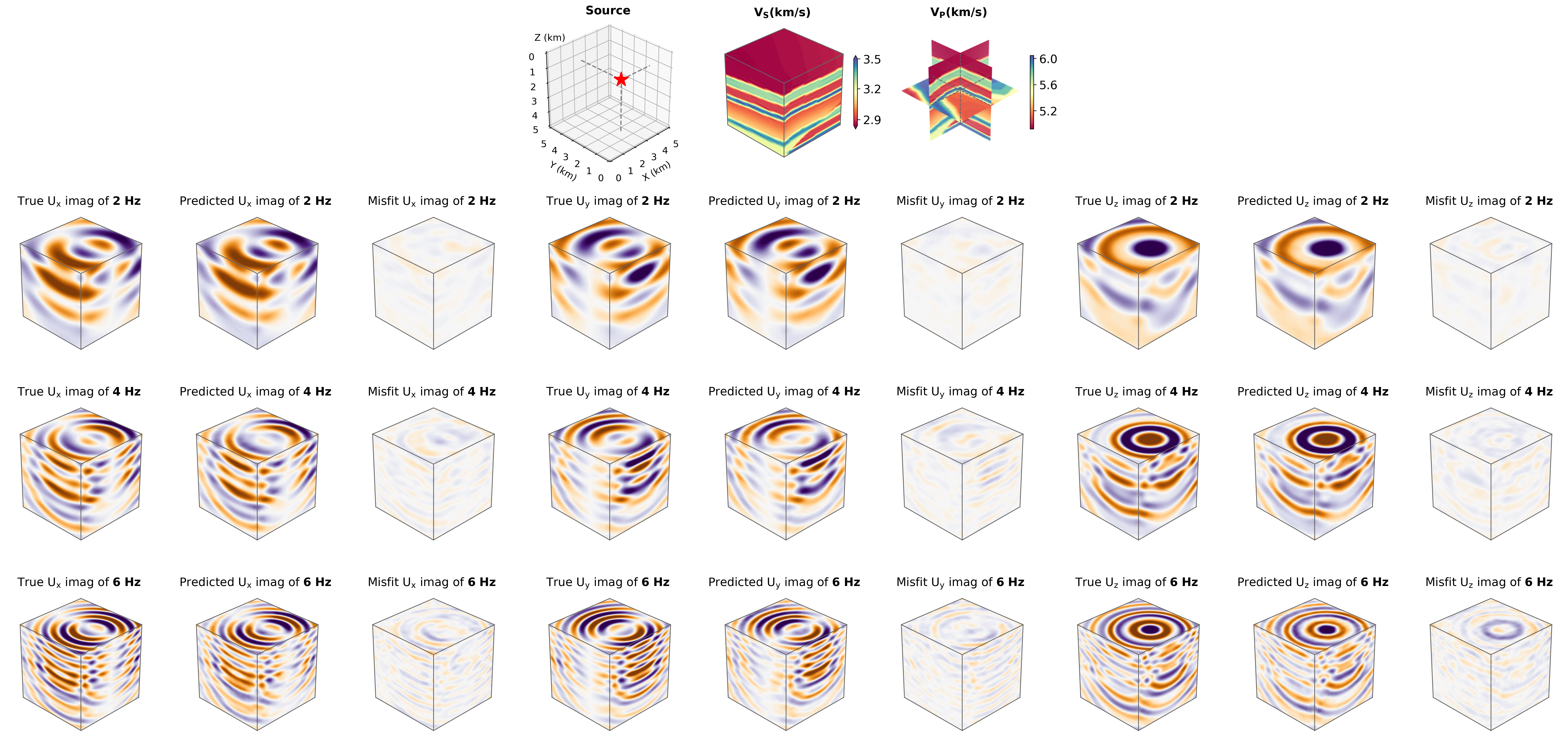}
    \caption{3D elastic wave modeling with U-NO evaluated in the frequency domain for an instance of random subpanels from the overthrust model. The first row displays the $V_P$ and $V_P/V_S$ models with the source location marked with a red star. The second, third, and fourth rows display the predicted results for the imaginary parts of displacement fields of 2 Hz, 4 Hz, and 6 Hz, respectively. The relative loss of the U-NO prediction is 0.110 for 2 Hz, 0.154 for 4 Hz, and 0.241 for 6 Hz.}
    \label{f3Doverthrustimag}
\end{figure*}

\begin{figure*}
    \centering
    \includegraphics[width=1\textwidth]{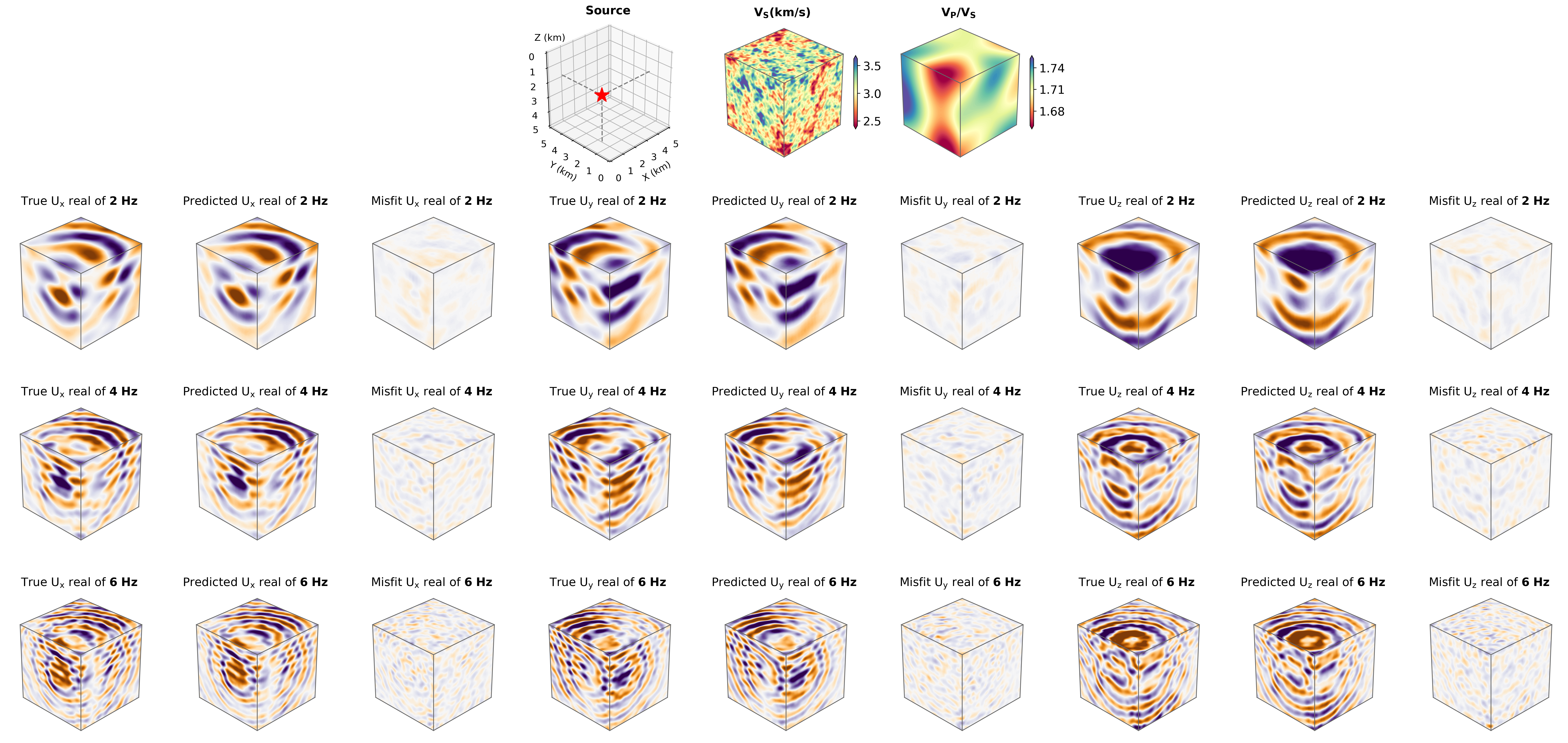}
    \caption{3D elastic wave modeling with U-NO evaluated in the frequency domain for an instance of random velocity fields. The first row displays the $V_P$ and $V_P/V_S$ models with the source location marked with a red star. The second, third, and fourth rows display the predicted results for the real parts of displacement fields of 2 Hz, 4 Hz, and 6 Hz, respectively. The relative loss of the U-NO prediction is 0.096 for 2 Hz, 0.149 for 4 Hz, and 0.238 for 6 Hz.}
    \label{f3Dreal}
\end{figure*}

\begin{figure*}
    \centering
    \includegraphics[width=1\textwidth]{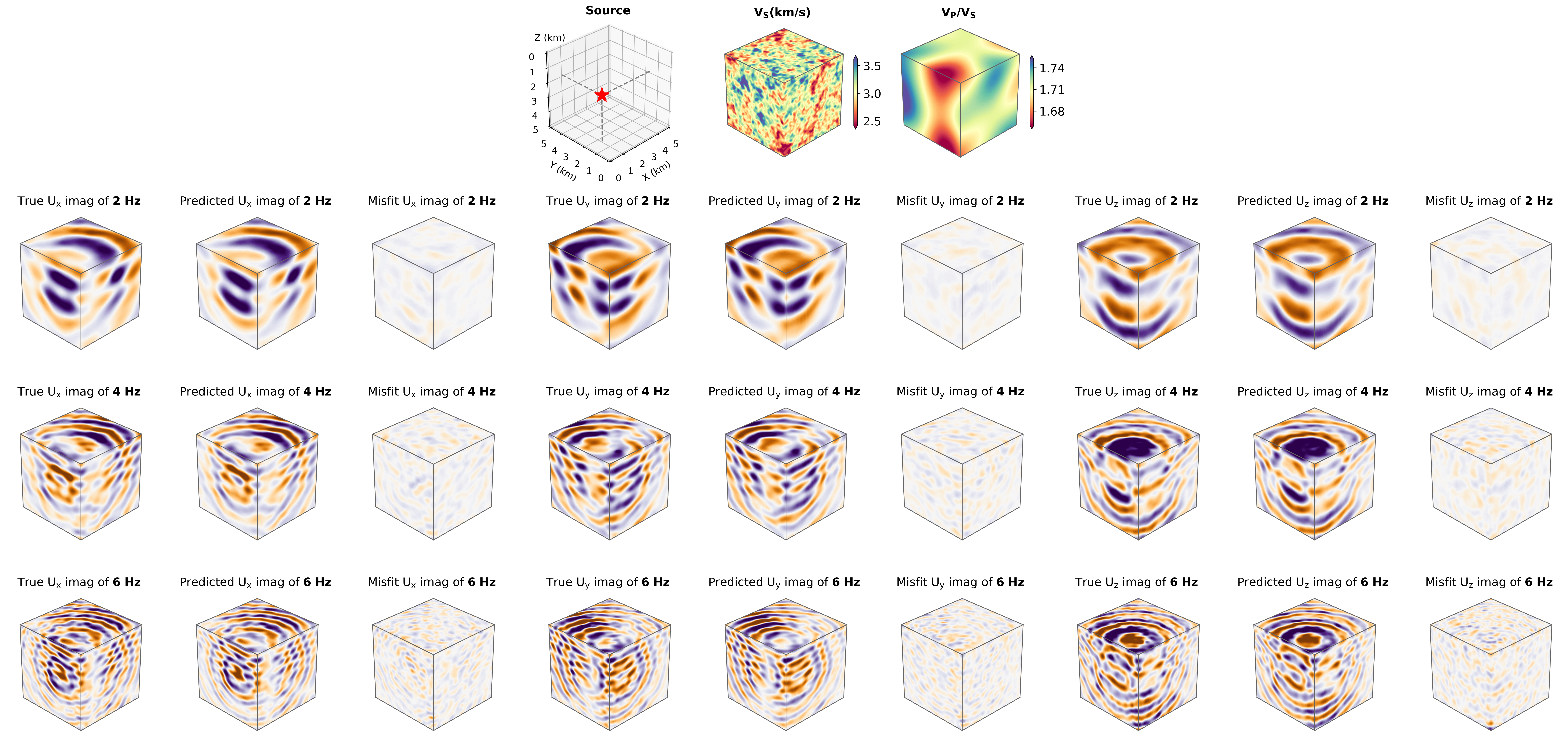}
    \caption{3D elastic wave modeling with U-NO evaluated in the frequency domain for an instance of random velocity fields. The first row displays the $V_P$ and $V_P/V_S$ models with the source location marked with a red star. The second, third, and fourth rows display the predicted results for the imaginary parts of displacement fields of 2 Hz, 4 Hz, and 6 Hz, respectively. The relative loss of the U-NO prediction is 0.096 for 2 Hz, 0.149 for 4 Hz, and 0.238 for 6 Hz.}
    \label{f3Dimag}
\end{figure*}

\begin{figure*}
    \centering
    \includegraphics[width=1\textwidth]{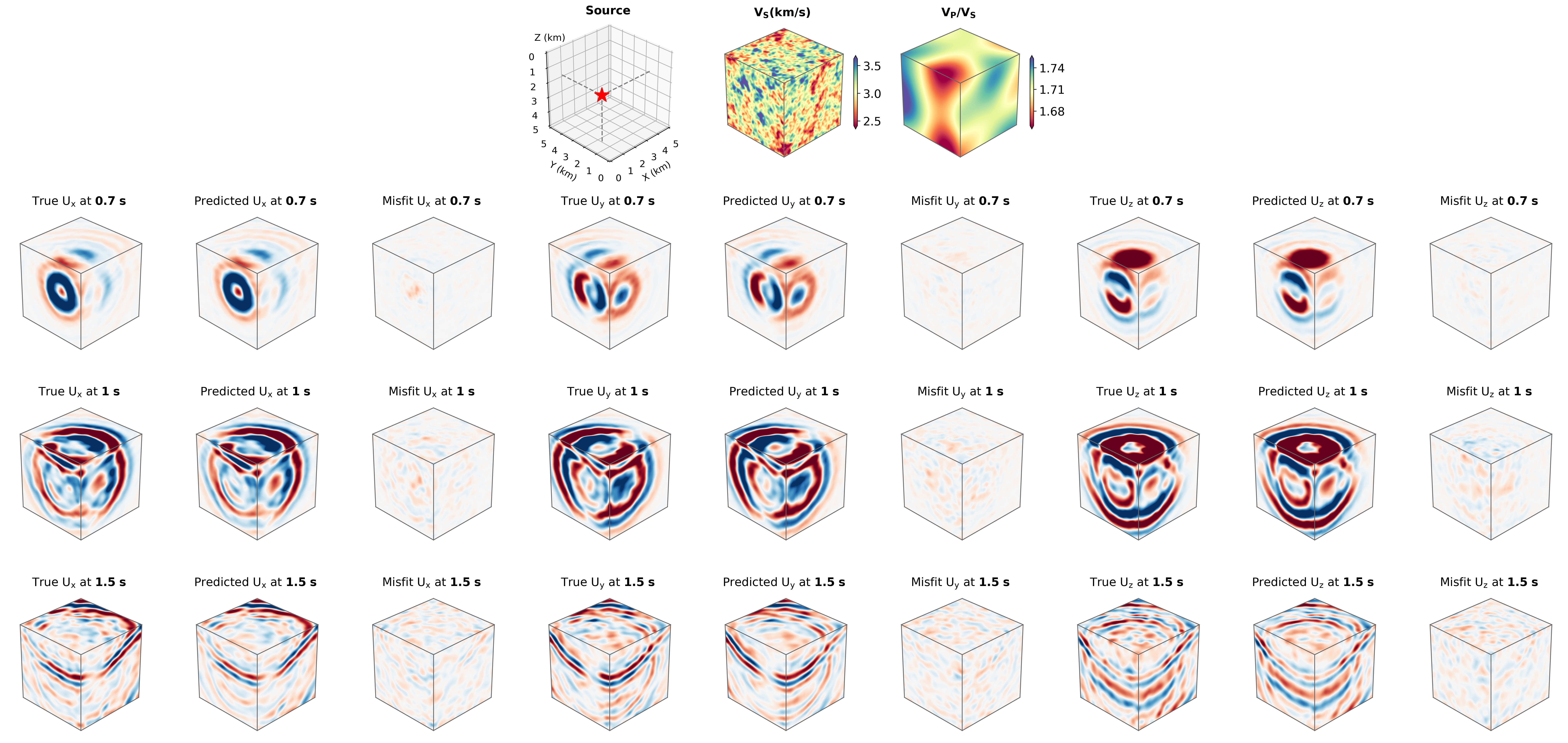}
    \caption{3D elastic wave modeling with U-NO evaluated in the time domain for an instance of random velocity fields. The first row displays the $V_P$ and $V_P/V_S$ models with the source location marked with a red star. The second, third, and fourth rows display the predicted results for displacement fields at 0.7 s, 1 s, and 1.5 s, respectively. The cross-correlation coefficient between the U-NO and SEM simulations is 0.989.}
    \label{t3D}
\end{figure*}
